%% file: main-tosem.tex
\documentclass[acmsmall, screen]{acmart}

\input{meta/lstcustom.tex}

\input{meta/algorithmstyle}
\input{meta/package.tex}

\input{meta/mintedcustom}

\newcommand{\blackding}[1]{\ding{\numexpr181+#1\relax}}
\newcommand{\whiteding}[1]{\ding{\numexpr171+#1\relax}}

\newcommand{\binxsight}{\text{BinXSight}}

\newcommand{\Anno}{\textsc{Anno}}
\newcommand{\AddEdge}{\textsc{AddEdge}}

\AtBeginDocument{%
  }

\setcopyright{acmlicensed}
\copyrightyear{2018}
\acmYear{2018}
\acmDOI{XXXXXXX.XXXXXXX}

\acmJournal{JACM}
\acmVolume{37}
\acmNumber{4}
\acmArticle{111}
\acmMonth{8}

\begin{document}

\title[BinXSight]{BinXSight: Fast Binary Program Summarization via Hierarchical Modular Decomposition}


\author{Zhijie Liu}
\email{rrocsha1793@gmail.com}
\affiliation{%
  \department{School of Information Science and Technology}
  \institution{ShanghaiTech University}
  \city{Shanghai}
  \country{China}
}

\author{Zhipeng Xue}
\authornote{Corresponding authors.}
\email{zhipengxue@zju.edu.cn}
\author{Zhipeng Gao}
\email{zhipeng.gao@zju.edu.cn}
\affiliation{%
  \institution{Zhejiang University}
  \city{Hangzhou}
  \country{China}
}

\author{Pengfei JING}
\email{philjin@foxmail.com}
\author{Zixiong Qin}
\email{zixiongqin@tencent.com}
\author{Sen Nie}
\email{snie@tencent.com}
\author{Shi Wu}
\email{shiwu@tencent.com}
\affiliation{%
  \institution{Tencent Security Keen Lab}
  \city{Shanghai}
  \country{China}
}

\author{Liang Feng Zhang}
\email{zhanglf@shanghaitech.edu.cn}
\affiliation{%
  \institution{ShanghaiTech University}
  \city{Shanghai}
  \country{China}
}

\author{Yutian Tang}
\authornotemark[2]
\email{yutian.tang@glasgow.ac.uk}
\affiliation{%
  \institution{University of Glasgow}
  \city{Glasgow}
  \country{United Kingdom}
}


\renewcommand{\shortauthors}{Liu et al.}

\input{secs/abstract}

\ccsdesc[500]{Security and privacy~Software reverse engineering}

\keywords{Binary, Program Summarization, LLM, Reverse Engineering}


\maketitle

\input{secs/introduction}

\input{secs/preliminary}

\input{secs/methodology}

\input{secs/evaluation}

\input{secs/discussion}

\input{secs/conclusion}

\normalem
\balance

\bibliographystyle{ACM-Reference-Format}
\bibliography{refs/refs}


\end{document}

%% file: meta/lstcustom.tex
\usepackage{color}
\usepackage{etoolbox}
\usepackage{listings}
\usepackage[T1]{fontenc}

\newcommand{\lstfontfamily}{\ttfamily}

\definecolor{darkviolet}{rgb}{0.5,0,0.4}
\definecolor{darkgreen}{rgb}{0,0.4,0.2} 
\definecolor{darkblue}{rgb}{0.1,0.1,0.9}
\definecolor{darkgrey}{rgb}{0.5,0.5,0.5}
\definecolor{lightblue}{rgb}{0.4,0.4,1}

\definecolor{stringColor}{rgb}{0.16,0.00,1.00}
\definecolor{annotationColor}{rgb}{0.39,0.39,0.39}
\definecolor{keywordColor}{rgb}{0.50,0.00,0.33}
\definecolor{commentColor}{rgb}{0.25,0.50,0.37}
\definecolor{javadocColor}{rgb}{0.25,0.37,0.75}
\definecolor{jTagColor}{rgb}{0.50,0.62,0.75}
\definecolor{eTagColor}{rgb}{0.50,0.62,0.75}
\definecolor{lineNumberColor}{rgb}{0.47,0.47,0.47}

\def\jTags{@author, @deprecated, @exception, @param, @return, @see, @serial, @serialData, @serialField, @since, @throws, @version}

\def\jAnnotations{
    classoffset=1,
    morekeywords={@Override, @Deperecated, @SuppressWarnings, @Retention, @Documented, @Target, @Inherited},
    keywordstyle=\color{annotationColor},
    classoffset=0
}

\def\eTags{FIXME, TODO, XXX}

\newrobustcmd{\markupJavadocs}[1]{%
\edef\mytok{\the\lst@token}%
\renewcommand*{\do}[1]{%
\ifdefstring{\mytok}{##1}%
{\color{jTagColor}\bfseries\listbreak}%
{}%
}%
{\color{javadocColor}%
\expandafter\docsvlist\expandafter{\jTags}%
\renewcommand*{\do}[1]{%
\ifdefstring{\mytok}{##1}%
{\color{eTagColor}\bfseries\listbreak}%
{}%
}%
\expandafter\docsvlist\expandafter{\eTags}%
#1}%
}%

\newrobustcmd{\markupComments}[1]{%
\edef\mytok{\the\lst@token}%
\renewcommand*{\do}[1]{%
\ifdefstring{\mytok}{##1}%
{\color{eTagColor}\bfseries\listbreak}%
{}%
}%
{\color{commentColor}%
\expandafter\docsvlist\expandafter{\eTags}#1}%
}%

\lstdefinestyle{eclipse}{
  basicstyle={\lstfontfamily},
  emphstyle=\bfseries,
  keywordstyle=\color{keywordColor}\bfseries,
  commentstyle=\markupComments,
  stringstyle=\color{stringColor},
  numberstyle=\color{lineNumberColor}\lstfontfamily,
  morecomment=[s][\markupJavadocs]{/**}{*/}, 
  showstringspaces=false,
  numbers=left,
}

\lstdefinestyle{black}{
  basicstyle=\small\lstfontfamily,
  numbers=left,
  columns=fullflexible,
  breaklines=true,
  mathescape=true,
  escapechar=\#,
  tabsize=4,
  frame=lines,
  showstringspaces=false
}

\lstdefinestyle{seminar}{
  basicstyle=\small\ttfamily,
  numbers=left,
  breaklines=true,
  mathescape=true,
  escapechar=\#,
  tabsize=4,
  showstringspaces=false
}

\expandafter\lstset\expandafter{\jAnnotations}

%% file: meta/algorithmstyle.tex
\usepackage{algorithm2e}
\usepackage{xcolor}

\SetAlFnt{\footnotesize}

\SetCommentSty{xCommentSty}

\LinesNumbered
\SetSideCommentRight
\DontPrintSemicolon
\RestyleAlgo{algoruled}

\usepackage{siunitx}
\usepackage[autolanguage]{numprint}

\newcommand*\np[2][z]{
\ifx z#1%
$\numprint{#2}$%
\else%
$\numprint[#1]{#2}$%
\fi\xspace%
}

\usepackage{fp}
\newcommand{\ShowAbsoluteNumber}[1]{%
\ifnum #1<10%
{\hspace*{0pt}#1}%
\else%
\ifnum #1<100%
{\hspace*{0pt}#1}%
\else%
\ifnum #1<1000%
{\hspace*{0pt}#1}%
\else%
{\numprint{#1}}%
\fi%
\fi%
\fi%
}

\newcommand{\ShowPercentage}[2]{%
\FPeval\percentage{round(#1/#2*100,0)}%
\FPeval\percentageOneDecimal{round(#1/#2*100,1)}%
\ifnum \percentage=0%
{\np[\%]{\FPprint{percentageOneDecimal}}}%
\else%
\ifnum \percentage<10%
{\np[\%]{\FPprint{percentageOneDecimal}}}%
\else%
{\np[\%]{\FPprint{percentageOneDecimal}}}%
\fi%
\fi%
\xspace
}
\newlength\BARSIZE  
\newcommand{\inlinechart}[2]{%
\FPeval{\BLACKBARSIZE}{#1/#2}\textcolor{black!80}{\rule{\BLACKBARSIZE\BARSIZE}{1.6ex}}%
\FPeval{\BLACKBARSIZE}{1 - (#1/#2)}\textcolor{black!10}{\rule{\BLACKBARSIZE\BARSIZE}{1.6ex}}%
}

\newcommand*\percent[3][v]{%
\ifx q#1%
    \np{#2}/\np{#3}(\ShowPercentage{#2}{#3})\else%
\ifx p#1%
    \np{#2}(\ShowPercentage{#2}{#3})\else%
\ifx c#1%
    \inlinechart{#2}{#3}%
\else%
    \np{#2}%
    \ifx r#1%
        /\np{#3}%
    \fi%
    \hspace*{0.5ex}(\ShowPercentage{#2}{#3}) %
    \inlinechart{#2}{#3}%
    \xspace
\fi\fi\fi%
}

%% file: meta/package.tex
\usepackage{hyperref}
\usepackage{color}
\usepackage{url}
\usepackage[most]{tcolorbox}
\usepackage{balance}
\usepackage{makecell}
\usepackage{verbatim}
\usepackage{ulem}
\usepackage{mdframed}
\usepackage{lipsum}
\usepackage{amsmath}
\usepackage{amssymb}
\usepackage{amsfonts}
\usepackage[font=small]{caption}
\usepackage{ifpdf}
\usepackage{graphicx}
\usepackage{wrapfig}
\usepackage{colortbl}
\usepackage{mathpartir}

\usepackage{pifont}

\usepackage[inkscapelatex=false]{svg}

\usepackage{algorithmic}
\usepackage{array}

\usepackage{caption}
\usepackage{multirow}
\usepackage{float}
\usepackage{mwe}
\newcommand{\code}[1]{\texttt{#1}}

\usepackage{url}
\usepackage{subfigure}
\usepackage{tikz}
\usepackage{xcolor}

\usepackage{booktabs,caption}  
\usepackage[flushleft]{threeparttable}

 \usepackage{enumitem}

\hypersetup{
 colorlinks=true,
 linkcolor=blue,
 citecolor=blue,
 filecolor=black,
 urlcolor=black,
 pdfauthor = {},
 pdftitle = {},
 pdfkeywords = {}
}

%% file: meta/mintedcustom.tex
\usepackage{fvextra}
\newcommand{\PYG}[2]{#2}

%% file: secs/abstract.tex
\begin{abstract}

Binary code summarization (BCS) is crucial in fields such as cybersecurity.
Particularly, program-level BCS can quickly provide a comprehensive overview of an entire stripped binary without in-depth manual reverse engineering. However, a binary often contains numerous functions with complex interactions, making it challenging to extract its functionality efficiently. Moreover, existing methods focus solely on function-level without considering contextual information, limiting their applicability to program-level analysis.
Towards comprehensive and fast program-level BCS, we propose \binxsight{}, an LLM-powered framework based on hierarchical modular decomposition.
Specifically, we first propose binary graph, a new data structure composed of four graphs, to comprehensively represent a binary.
Then, based on the binary graph, we hierarchically decompose functions into modules and select the most significant ones.
Finally, \binxsight{} generates a program-level summary from these module summaries.
We evaluate \binxsight{} on stripped binaries across various sizes, fields, platforms, and architectures.
Results show that \binxsight{} outperforms baselines by 23.18\% to 46.99\%, and summarizes a binary with about 3K functions in 24.3 seconds on average.
In addition, a user study with 27 experts shows that \binxsight{}'s high-quality summaries help users quickly understand the whole binary.
Our case analysis on recent real-world malware further demonstrates its practical utility.
\looseness=-1

\end{abstract}

%% file: secs/introduction.tex
\section{Introduction}\label{sec:introduction}

Binary code summarization (BCS) is a crucial research area in the fields of reverse engineering, binary analysis, and cybersecurity~\cite{symgen:2025, ye2023cp, jin2022symlm, gao2021lightweight, patrick2023xfl, jin2023binary, jin2024simllm}. It aims to generate concise and human-readable summaries for (stripped) binary code, which can help security analysts efficiently comprehend the functionality, eliminating the need for a comprehensive and time-consuming reverse engineering process.
Among these techniques, \textit{program-level BCS} stands out as a powerful
technique as it can quickly provide a comprehensive overview of the functionality and behaviors of an entire binary (program). Reverse engineers can quickly understand the functionality of a binary without having to sift through thousands of functions.
\looseness=-1

Unfortunately, despite the significance of program-level BCS, none of the previous works could achieve satisfying results. Existing studies~\cite{zhu2025misum, symgen:2025, jin2023binary, ye2023cp, lu2024malsight} mainly focus on function-level.
As a result, these methods still require manual inspection of all functions to understand the binary's functionality, and inferring the program summary from them suffers from the following limitation.
\looseness=-1

Typically, a binary may contain \textit{tens of thousands} of functions. Its functionality comprises the intricate inter-function dependencies, making it challenging to aggregate function results into a coherent program overview. In addition, existing function-level techniques~\cite{symgen:2025, zhu2025misum, jin2023binary} often neglect the calling context of a function to ensure efficiency. However, failing to consider such context can restrict the actual functions' behaviors accessible to the summary models~\cite{hu2024degpt, jin2022symlm}, leading to significant information loss during the summarization process from functions to program.

Inspired by the modular design principle in software development~\cite{gallobject, tang2021xdebloat}, we can divide a binary into multiple functional modules. Each module, composed of function interactions, has rich behavioral semantics and corresponds to a complete feature for users~\cite{tang2021xdebloat, liu2023enhancing}, allowing us to extract functionality from them. In addition, the context can be more easily considered at module scope than at program scope. Nevertheless, adopting modules for program summarization while maintaining efficiency still poses several \textbf{challenges}.
\looseness=-1

\noindent $\bullet$ \textbf{C1:} \textbf{Comprehensive Representation for Binary.} One binary can be composed of numerous functions with complex interactions and dependencies, not limited to calling relationships. Therefore, representing it to delineate comprehensive functional dependencies is challenging.

\noindent $\bullet$ \textbf{\textbf{C2:}} \textbf{Decomposition into Functional Modules.} Decomposing a complex binary representation into modules requires complete behavioral semantic retention and compatibility for semantics extraction, making effective module generation a formidable challenge.

\noindent $\bullet$ \textbf{\textbf{C3:}} \textbf{Core Behavioral Semantics Extraction from Modules.} A binary often exhibits varied behaviors from its modules, including those contributing to core behaviors and those introducing noise (e.g., error handling). Thus, extracting core behavioral semantics from all modules for comprehensive and fast summarization remains challenging.\looseness=-1

To fill the above gaps, we propose \binxsight{}, a novel LLM-powered framework designed to quickly generate program summaries for stripped binaries via hierarchical modular decomposition.
Specifically, \binxsight{} involves two stages: binary graph construction (Sec. \ref{sec:binarygraphconstruction}) and hierarchical summarization (Sec. \ref{sec:hierarchicalsummarization}).
In the first stage, we present a binary into a new form of binary graph, a representation that contains comprehensive multi-perspective semantics of program behaviors of the binary and forms a clustered distribution of these behaviors.
In the second stage, we first decompose the binary graph into modules hierarchically, ensuring each module preserves rich behavioral semantics and is suitable for inference through large language model (LLM)~\cite{openai, gpt-4o-mini, liu2024deepseek}.
Next, \binxsight{} selects high-value modules among all the generated ones and leverages the MapReduce strategy to summarize them for extracting the core behavioral semantics. Eventually, we reconstruct all summarized modules with their hierarchy information to LLM for generating the final comprehensive program summary.
\looseness=-1

We perform experiments on a test dataset with stripped binaries across various sizes, fields, platforms, and architectures.
\binxsight{} delivers effective performance, achieving 0.6137 in SIMILARITY and significantly outperforming baselines by 23.18\% to 46.99\%.
Meanwhile, it shows remarkable efficiency. On average, it processes a binary with about 3,429 functions in just 24.3 seconds, and even for a large binary with 59,173 functions, the processing time remains low at 3.8 minutes.
In addition, we invited 27 experts to evaluate the quality of its summaries. The result indicates that \binxsight{} can generate high-quality summaries that help reverse engineers quickly understand the whole binary in the first place.
Our case analysis further shows that \binxsight{} helps security vendors identify suspicious and malicious behaviors for recent active real-world malware.
\looseness=-1

To summarize, our paper makes following contributions:

\noindent $\bullet$ To the best of our knowledge, we are the \textit{\textbf{first}} to introduce modules for program summarization for stripped binaries, enabling comprehensive and fast binary program understanding for reverse engineers, without manual in-depth analysis;

\noindent $\bullet$ We propose \binxsight{}, a novel LLM-powered framework for program-level BCS. By leveraging binary graph construction and hierarchical summarization, it can comprehensively summarize binaries while maintaining efficiency;

\noindent $\bullet$ We evaluate \binxsight{} on diverse stripped binaries. \binxsight{} achieves effective performance and significantly outperforms baselines by 23.18\% to 46.99\%. Meanwhile, it shows efficiency in program summarization. Our user survey also indicates its usefulness in aiding reverse engineers in understanding unknown binaries, and our case analysis shows its practical utility in real-world malware analysis.

%% file: secs/preliminary.tex
\section{Preliminary, Related Work, and Motivation}\label{sec:preliminary}

\begin{table}[t]
\centering
\caption{Granularity and output of representative binary reverse engineering approaches.}
\label{tab:related-comparison}
\scalebox{0.75}{\begin{tabular}{lll}
\toprule
\textbf{Method} & \textbf{Granularity} & \textbf{Output} \\
\midrule

SymLM~\cite{jin2022symlm}       & Function & Predict function names \\
DeGPT~\cite{hu2024degpt}        & Function & Restructure decompiled code \\
ReSym~\cite{xie2024resym}       & Function & Recover variable names and data structures \\
GenNm~\cite{xu2025unleashing}   & Function & Recover variable names \\
SymGen~\cite{symgen:2025}       & Function & Predict function names \\

\midrule

CP-BCS~\cite{ye2023cp}          & Function & Per-function summarization \\
BinT5~\cite{al2023extending}    & Function & Per-function summarization \\
MALSIGHT~\cite{lu2024malsight}  & Function & Per-function summarization \\
MiSum~\cite{zhu2025misum}       & Function & Per-function summarization \\

\midrule

\textbf{\binxsight{}}    & \textbf{Module / Program} & \textbf{Program-level description from module summaries} \\
\bottomrule
\end{tabular}}
\end{table}

\subsection{Binary Reverse Engineering}\label{sec:binaryreverseengineering}

Binary reverse engineering analyzes binaries to recover their functionalities without source code access. This process is particularly challenging when stripping is applied to remove semantically meaningful debugging symbols from binaries, such as function names and type information. To mitigate these challenges, many techniques~\cite{symgen:2025, hu2024degpt, xie2024resym, dramko2026idioms} have been proposed to recover the semantic information.\looseness=-1

Most of the existing works leverage deep learning (DL) for symbol recovery tasks, e.g., function name prediction~\cite{symgen:2025, jin2022symlm, gao2021lightweight, david2020neural}, variable name/type recovery~\cite{hu2024degpt, xie2024resym, chen2022augmenting, pal2024len, zhang2021osprey, xu2025unleashing}, and so forth. For example, DeGPT~\cite{hu2024degpt} prompts ChatGPT~\cite{openai} to construct a three-role mechanism to perform pseudocode optimization, including code structures, variable names, and comments. SymGen~\cite{symgen:2025} employs a domain adaptation strategy and fine-tunes LLM with LoRA~\cite{hu2022lora} for generalizable function name prediction.
\looseness=-1

Another line of research focuses on generating more high-level and human-readable semantic descriptions of binary functions~\cite{zhu2025misum, ye2023cp, al2023extending, jin2023binary, lu2024malsight}. For instance, CP-BCS~\cite{ye2023cp} employs the control flow graph and pseudo code enriched with expert knowledge to learn comprehensive binary function behaviors. MiSum~\cite{zhu2025misum} gathers both assembly code and pseudocode into a multi-modality heterogeneous code graph with an intent-aware generator for summarizing.

However, as summarized in Table~\ref{tab:related-comparison}, these methods only reach function-level granularity, so understanding an entire binary still requires manual inspection of all functions.

\subsection{Definition of Binary Program Summary}\label{sec:definitionofbps}

A well-crafted ground-truth program summary is essential as it serves as a benchmark for evaluating effectiveness.
In this study, we define a comprehensive ground-truth summary as $\text{\textsf{Summary}} = \langle I, F, B, C\rangle$ that includes the following essential components to ensure utility:

\begin{itemize}[leftmargin = *]
    \item \textbf{Introduction ($\mathbf{I}$).} A brief statement outlining the primary functionality and intended use of the binary;

    \item \textbf{Functionality ($\mathbf{F}$).} A detailed description of what the tasks presented by the binary (e.g., system manipulation);

    \item \textbf{Behaviors ($\mathbf{B}$).} A detailed description of how the actions exhibited by the binary (e.g., remote screen control);

    \item \textbf{Conclusion ($\mathbf{C}$).} A summary that encapsulates the main points discussed above.
\end{itemize}

\noindent We provide an example as below:
\looseness=-1

\textbf{gh0st~\cite{gh0st}.} 
"\textit{$\mathbf{I}$: The binary program, a sophisticated remote access client, enables attackers to remotely control and monitor compromised machines, facilitating unauthorized access to sensitive information.
$\mathbf{B}$: It is adept at data theft, targeting sensitive system information, personal and financial data, and intellectual property. Additionally, the program supports remote control and monitoring of various system behaviors, including remote screen control and capture, process and window enumeration, registry and service monitoring, file manipulation, multimedia recording, and keyboard input logging.
$\mathbf{F}$: These capabilities facilitate real-time surveillance of the victim's activities, making it a potent tool for unauthorized data collection and system manipulation.
$\mathbf{C}$: Overall, the binary program allows for the compromised control and monitoring of machines, and excels at stealthy and malicious data theft, making it a significant threat in the realm of cybersecurity.}"

\subsection{Motivation}\label{sec:motivation}

Program-level BCS aims to provide a comprehensive overview of a binary, yet, as discussed in Sec.~\ref{sec:introduction}, existing function-level methods struggle to extract such functionality while maintaining efficiency.
Inspired by the modular design principle~\cite{gallobject, yang2022modx, liu2023enhancing, tang2021xdebloat} in software development, each functional module in a program is highly cohesive (i.e., high modularity) and represents a complete and distinct feature for users~\cite{berger2015feature, lettner2015feature, behringer2017peopl}. That is, functions in a module target a specific program behavior (i.e., feature), and the functionality of a binary resides in these features.
In addition, modules are loosely coupled, which means they can be considered approximately isolated from each other. These two properties create an opportunity to focus solely on the modules, allowing for efficiency while ensuring the core behavioral semantics are captured, as program-level issues are converted into smaller, parallelizable module-level tasks. \textit{Thus, the entire program-level BCS is transformed into how to decompose and generate modules from a complex binary and leverage these modules to extract the functionality of the program.}
Based on it, we propose \binxsight{} for comprehensive and fast program-level BCS. \looseness=-1

\begin{figure*}[t]
\centering
\includegraphics[width=1.0\textwidth]{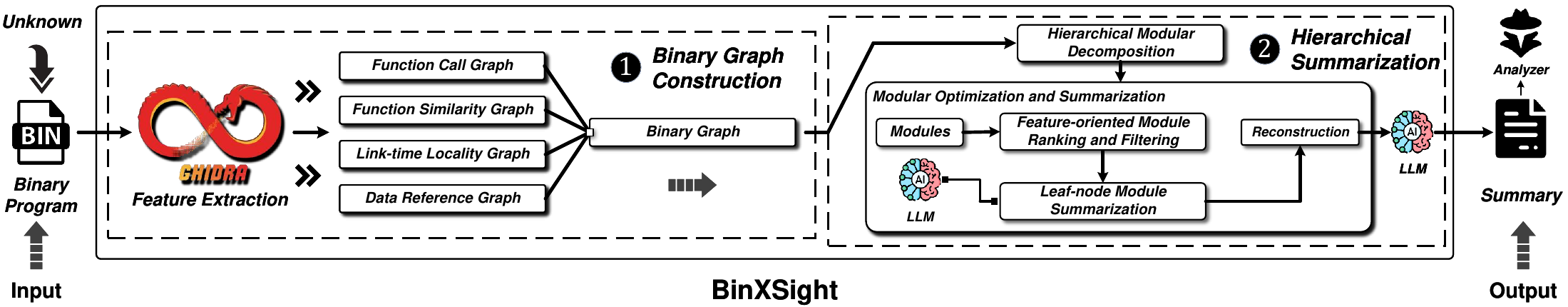}
\caption{The workflow of \binxsight{}.}
\label{fig:workflow}
\end{figure*}

%% file: secs/methodology.tex
\section{Methodology}\label{sec:methodology}

As shown in Fig. \ref{fig:workflow}, \binxsight{} includes two phases: \blackding{1} binary graph construction (Sec. \ref{sec:binarygraphconstruction}), and \blackding{2} hierarchical summarization (Sec. \ref{sec:hierarchicalsummarization}).
Specifically, to summarize a given binary, \blackding{1} \binxsight{} extracts decompiled C-like pseudocode functions, control flow information, and data references through Ghidra~\cite{ghidra}. Accordingly, it constructs a heterogeneous binary graph for the binary based on the three features.
Then, \blackding{2} \binxsight{} leverages \textit{Hierarchical Modular Decomposition} to decompose the binary graph into functional modules hierarchically. Next, it uses \textit{Modular Optimization and Summarization} to select high-value modules (i.e., more significant to core behaviors) and summarize them through LLM.
Eventually, after reconstructing the module summaries with the hierarchy structure, \binxsight{} gathers all of them to LLM to generate the program summary, describing the whole functionality and behaviors from a comprehensive perspective.
\looseness=-1

\subsection{Binary Graph Construction}\label{sec:binarygraphconstruction}

\noindent \textbf{Challenge: How to comprehensively represent a binary?} The key idea behind \binxsight{} is to generate the program summary through functional modules, where each module represents a group of functions that indicate a specific program behavior of the binary.
\textbf{C1:} Previous works~\cite{liu2023enhancing, tang2021xdebloat} primarily rely on function call graph (FCG) for functional module generation.
However, the functionality of a binary is not solely presented by the calling relationships.
They also inherently encompass other interactions and relationships, such as function semantics and data dependencies.
The lack of involvement makes the representation incomplete, affecting later module generation.
\looseness=-1

\smallskip
\noindent \textbf{Solution: Construct the Binary Graph.} \textbf{S1:} We propose \textit{binary graph}, a heterogeneous representation composed of four graphs capturing complementary perspectives: function call graph models the global behavioral distribution via invocations; function similarity graph connects behaviorally similar functions to densify functional semantics; link-time locality graph captures function order within compile units and the boundary between user code and third-party libraries (TPL); and data reference graph reflects implicit data dependencies via shared data accesses. The representation naturally forms a comprehensively clustered distribution of a binary's behaviors, providing an effective data structure for module generation (Sec.~\ref{sec:hierarchicalmodulardecomposition}).
\looseness=-1

\subsubsection{Feature Extraction}\label{sec:featureextraction}

We leverage Ghidra~\cite{ghidra} to analyze the given binary:

\smallskip
\noindent \textbf{Pseudo Function.} C-like pseudocode functions are decompiled from the binary functions, and close to the code in high-level languages of C/C++. This makes it easy for LLM to understand~\cite{jin2023binary, symgen:2025}. We also extract the relative virtual address (\textit{rva}) of each function as the ordinal number, indicating the link-time locality in the binary~\cite{lfa, jiang2024binaryai}.

\smallskip
\noindent \textbf{Control Flow.} This information is crucial for modeling program behaviors among binary functions~\cite{tang2021xdebloat, liu2023enhancing}, illustrating how one function calls another and the sequence of execution. We extract the control flow of the calling relationship with XREF~\cite{ghidra}.

\smallskip
\noindent \textbf{Data Reference.} Data reference captures how functions access the same data elements, reflecting the implicit data dependencies and behaviors within a program.
In a binary, global variables, static variables, and string literals are allocated in the static memory regions (e.g., \code{.data}, \code{.bss}, and \code{.rodata} sections in ELF), ensuring their persistence throughout execution and is accessible across functions.
We extract these relationships with XREF.
\looseness=-1

\subsubsection{Graph Construction}\label{sec:graphconstruction}

\quad

\smallskip
\noindent \textbf{Function Call Graph (FCG).} Intuitively, a large number of inter-calling relationships among a group of functions indicates that they are likely to contribute to a specific program behavior~\cite{gallobject, liu2023enhancing, tang2021xdebloat}. Groups are decoupled through minimal calling relationships (e.g., interface)~\cite{gallobject, liu2023enhancing, tang2021xdebloat}. \textit{\textit{Thereby, from a global perspective, FCG characterizes the initial behavioral distribution of the binary.}} 

We construct FCG from control flow.
Formally, FCG is a simple directed graph with loops $G_{\text{FCG}} = \{V_{f}, E_{c}\}$. $V_{f}$ represents the pseudo function vertices, and $E_{c}$ represents the calling relationship edges. For a function $v_i \in V_{f}$ invokes another function $v_j \in V_{f}$ multiple times, we only add one edge $e_{ij}$ from $v_i$ to $v_j$ into $E_{c}$, avoiding excessive local connection density within a specific program behavior~\cite{liu2023enhancing}. We further leverage a widely-used value-set analysis (VSA) framework~\cite{binabsinspector} to resolve the indirect calls, improving the integrity of FCG. \looseness=-1

\begin{wrapfigure}{r}{0.6\textwidth}
\centering
\includegraphics[width=0.6\textwidth]{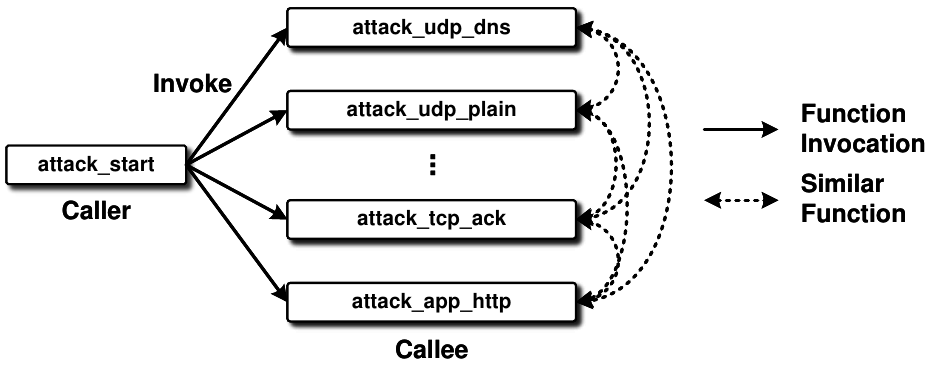}
\caption{The function \code{attack\_start} indirectly invokes specific attack functions, forming the complete attack chain.}
\label{fig:attack-func-sim}
\end{wrapfigure}

\smallskip
\noindent \textbf{Function Similarity Graph (FSG).} In a binary, multiple functions can exhibit similar behaviors. Aggregating them reveals functional intent from a semantic perspective~\cite{wang2024clap}, enhancing local connection density to better partition modules later.
For example, in the Mirai~\cite{antonakakis2017understanding} botnet, the attack driver function \code{attack\_start} (see Fig. \ref{fig:attack-func-sim}) dispatches multiple attack functions, e.g., \code{attack\_udp\_dns}, via indirect calls (Fig. \ref{fig:attack-func-sim}'s \code{Invoke}) to launch DDoS attacks. 
These attack functions, behaviorally similar to each other (e.g., TCP/UDP packet sending), have no control or data dependency among them. Based on FCG, the driver and attack functions fall into separate modules, as the invocations from the attack functions to their callees are more dense than the driver's dispatching them. Relying solely on FCG, LLM cannot see the complete attack chain, resulting in fragmented insights into these network behaviors.
Instead, connecting these attack functions with function similarity edges (Fig. \ref{fig:attack-func-sim}'s similar function) can enhance the local connection density among them and the attack driver. This allows the complete attack chain to be grouped into one module, enabling LLM to observe full attack behaviors. \textit{Through FSG, such implicit behavioral relationships among functions can be discovered, contributing to generating a more behaviorally comprehensive functional module.} \looseness=-1

Specifically, FSG is defined as a simple undirected graph $G_{\text{FSG}} = \{V_{f}, E_{s}\}$ where $E_{s}$ represents the function similarity edges. We use the state-of-the-art KEENHash embedding model~\cite{keenhash:2025} to encode each pseudo function $v_i \in V_f$. These embeddings represent the semantics and can be used to calculate the cosine similarity between them. Thus, we compute an upper triangular matrix of similarity, where each element $s_{ij}$ represents the similarity score between pairs of functions of $v_i$ and $v_j$ ($i \neq j$). 
If $s_{ij} \geq \text{a threshold } t$, then we add an edge $e_{ij}$ into $E_{s}$ between $v_i$ and $v_j$.
We set $t = 0.75$ based on our experience.
\looseness=-1

\smallskip
\noindent \textbf{Link-time Locality Graph (LLG).} Functions located in the same source file (i.e., compile unit) often contribute to a sub-behavior of a program behavior~\cite{lang2024delink}. Connecting them can further enhance the functional intent, increasing their density within the graph.
For example, the functions in \code{attack.c} of \code{attack\_start} (Fig.~\ref{fig:attack-func-sim}) contribute to dispatching specific attacks from other files, like \code{attack\_app\_http} in \code{attack\_app.c}. In addition, the link-time localities of the binary functions compiled from the same source file are rendered continuous in the binary~\cite{jiang2024binaryai}. \textit{\textit{Thus, we can derive that functions located near each other likely support the same sub-behavior, implying locality dependencies.}} Furthermore, with the widespread reuse of TPLs~\cite{yang2022modx, jiang2024binaryai}, a binary often encompasses both user-developed and TPL code. They may be located interleavingly within the code section. Their boundaries need to be delineated, as the behavioral relationships between them do not depend on their localities.
\looseness=-1

\begin{figure}[t]
\centering
\scalebox{0.76}{%
\begin{minipage}{1.3\linewidth}
\begin{mathpar}

\inferrule*[right=\textsc{User-Locality}]
{ \neg \Anno(v_i) \;\land\; \neg \Anno(v_j) }
{ \AddEdge(v_i, v_j) }
\and
\inferrule*[right=\textsc{Cross-Boundary-1}]
{ (\Anno(v_i) \land \neg \Anno(v_j)) \;\lor\; (\neg \Anno(v_i) \land \Anno(v_j)) }
{ \neg \AddEdge(v_i, v_j) }

\\

\inferrule*[right=\textsc{Cross-Boundary-2}]
{ \Anno(v_i)=\langle p_i,f_i\rangle \\ \Anno(v_j)=\langle p_j,f_j\rangle \\ \langle p_i,f_i\rangle\neq\langle p_j,f_j\rangle }
{ \neg \AddEdge(v_i, v_j) }
\and
\inferrule*[right=\textsc{TPL-Locality}]
{ \Anno(v_i)=\langle p_i,f_i\rangle \\ \Anno(v_j)=\langle p_j,f_j\rangle \\ \langle p_i,f_i\rangle = \langle p_j,f_j\rangle }
{ \AddEdge(v_i, v_j) }

\end{mathpar}

\smallskip
$v_i, v_j$ denote two consecutive functions, ordered by their \textit{rva} such that
\[
rva(v_i) < rva(v_j)
\;\land\;
\forall v_k \in V_f \setminus \{v_i, v_j\},\;
\neg (rva(v_i) < rva(v_k) < rva(v_j)).
\]
$\Anno(v_i)$ returns the annotation of $v_i$, where $p_i$ is the project name to which $v_i$ belongs, and $f_i$ is the source file in $p_i$, having $v_i$. $\AddEdge(v_i, v_j)$ indicates a locality edge added from $v_i$ to $v_j$.
\end{minipage}
}
\caption{Inference rules defining the relation $\textsc{AddEdge}(v_i, v_j)$ 
for constructing Link-time Locality Graph.}
\label{fig:llg-rules}
\end{figure}

LLG is a simple directed graph $G_{\text{LLG}} = \{V_{f}, E_{l}\}$ where $E_{l}$ is the function (link-time) locality edges. We sort functions by ascending \textit{rva} and use BinaryAI~\cite{jiang2024binaryai}, the state-of-the-art software composition analysis tool, to detect TPLs. Functions $v_i \in V_f$ identified as belonging to TPLs are annotated with the project names of the TPLs and the source files. We then linearly scan all consecutive function pairs <$v_i$, $v_j$> by \textit{rva}. Whether adding edge $e_{ij}$ from $v_i$ to $v_j$ follows the four rules in Fig. \ref{fig:llg-rules}. Their descriptions are listed below:
\looseness=-1

\begin{itemize}[leftmargin = *]
    \item \textsc{User-Locality}: If both functions have no annotation, then we add $e_{ij}$, indicating the locality within user-developed code;

    \item \textsc{Cross-Boundary-1}: If only one function is annotated, then no edge is added, indicating the boundary between user-developed code and TPL code;

    \item \textsc{Cross-Boundary-2}: If both functions are annotated but their <project name, source file> are not equivalent, then no edge is added, indicating the boundary between two compile units;

    \item \textsc{TPL-Locality}: If both functions are annotated and their <project name, source file> are equivalent, then we add $e_{ij}$, indicating the locality within the same compile unit.
\end{itemize}

\smallskip
\noindent \textbf{Data Reference Graph (DRG).} Functions accessing the same data likely contribute to the same program behavior. \textit{\textit{Thus, DRG can complement the function interactions based on implicit data dependencies.}} For example, Mirai uses global variables to pass the attack function pointers, ultimately reaching \code{attack\_start}.
We leverage data reference to construct DRG. DRG is defined as a simple directed graph $G_{\text{DRG}} = \{V_{a}, E_{d}\}$. $V_{a} = V_{f} \cup V_{d}$ where $V_{d}$ represents accessed data vertices; $E_{d}$ represents the data reference edges. If a function $v_i \in V_{f}$ accesses a data object $v_j \in {V_{d}}$, we add an edge $e_{ij}$ from $v_i$ to $v_j$ into $E_{d}$.
\looseness=-1

\smallskip
\noindent \textbf{Binary Graph.} The binary graph combines the above four graphs into a directed multigraph $G = \{V, E\}$ where $V = V_f \cup V_a$ and $E = E_c \cup E_s \cup E_l \cup E_d$. Before construction, FSG is first transformed into a symmetric directed graph to match this directedness.

\subsection{Hierarchical Summarization}\label{sec:hierarchicalsummarization}

\noindent \textbf{Challenge: How to comprehensively and quickly capture functionality from complex binary graph?}
We decompose binary graph into functional modules and extract the functionality from them for program summarization.
\noindent \textbf{C2:} However, decomposing an extremely large and complex binary graph requires that the resulting modules retain complete behavioral semantics while staying within the LLM's token limits for effective inference. Consequently, effective module generation presents a formidable challenge.

\begin{wrapfigure}{r}{0.53\textwidth}
\centering
\includegraphics[width=0.53\textwidth]{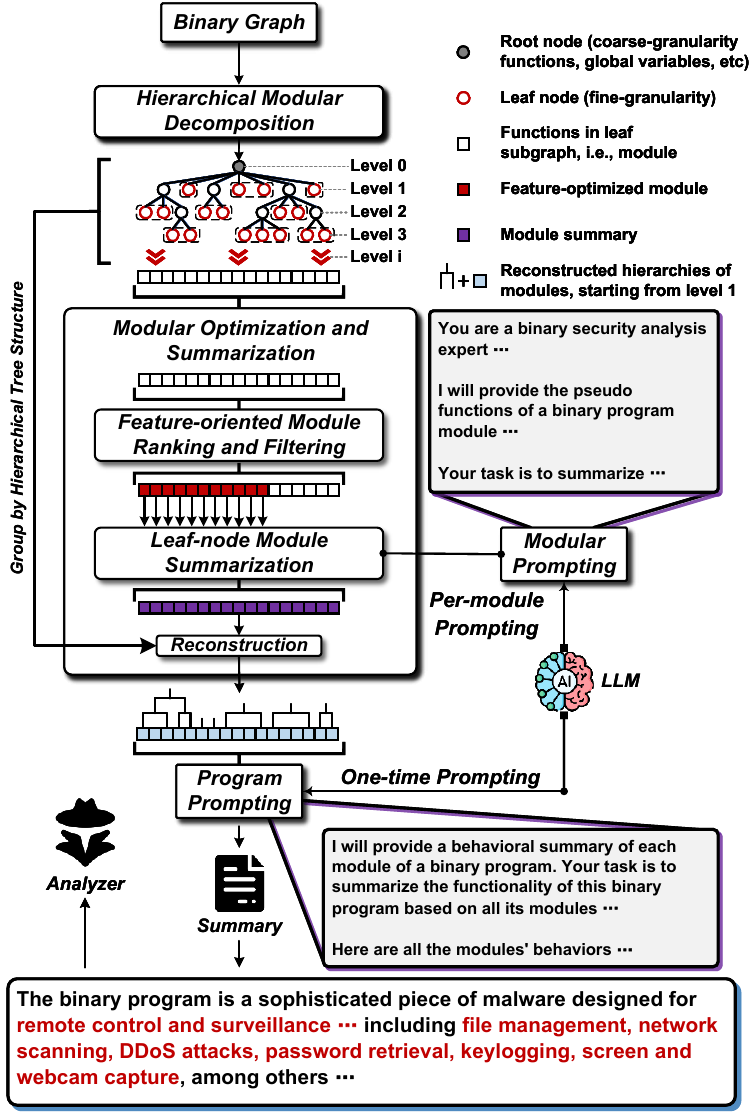}
\caption{The workflow of Hierarchical Summarization.}
\label{fig:modular-sum}
\end{wrapfigure}

\textbf{C3:} In addition, we extract the functionality from module summaries, nevertheless, a large binary can also result in numerous modules and exhibit various behaviors. These behaviors can contain both core behavioral semantics and noise (e.g., error handling). Directly summarizing all modules can introduce invalid information, affecting the quality of the program summary. It can also impact the summarization efficiency and economic costs. Moreover, summarizing a module is non-trivial, as it needs to consider not only the inner function semantics but also their contextual interactions to present complete behaviors. Therefore, extracting the core behavioral semantics from modules to achieve comprehensive and fast program summarization remains challenging. \looseness=-1

\smallskip
\noindent \textbf{Solution: Leverage Hierarchical Summarization.} We design \textit{hierarchical summarization} strategy to decompose binary graph and select high-value modules for module and program summarization in a MapReduce manner~\cite{dean2008mapreduce}.

As shown in Fig. \ref{fig:modular-sum}, hierarchical summarization includes three steps: \blackding{1} hierarchical modular decomposition (Sec. \ref{sec:hierarchicalmodulardecomposition}); \blackding{2} modular optimization and summarization (Sec. \ref{sec:modularoptimizationandsummarization}); and, \blackding{3} program prompting (Sec. \ref{sec:programprompting}).
\textbf{S2:} In particular, \blackding{1} \binxsight{} leverages \textit{Hierarchical Modular Decomposition} to decompose a binary graph into functional modules hierarchically, in a tree structure. It provides fine-granularity groups of functions (i.e., modules, leaf nodes in Fig. \ref{fig:modular-sum}), preserving rich behavioral semantics and suitable for LLM to summarize.
\textbf{S3:} Next, \blackding{2} in \textit{Modular Optimization and Summarization}, \binxsight{} heuristically ranks the modules on their features, filtering high-value ones through \textit{Feature-oriented Module Ranking and Filtering}. Then, these modules are sent to LLM in parallel (\textit{Map Phase}). Each module summarization is in a one-conversation manner with considering contextual information. After receiving these summaries, they are grouped by their original hierarchy using \textit{Reconstruction}.
Eventually, \blackding{3} \binxsight{} sends the reconstructed module summaries to LLM to generate the program summary (\textit{Reduce Phase}).
\looseness=-1

\subsubsection{Hierarchical Modular Decomposition}\label{sec:hierarchicalmodulardecomposition}

To be suitable for summarization, the generated modules need to meet two conditions: (1) each module represents a complete and distinct feature for users; and (2) the generated module is suitable for the token limits (i.e., context window) of LLM. 

For the first condition, we leverage Louvain community detection algorithm~\cite{ghosh2018distributed} to partition binary graph, choosen for two reasons: (1) Louvain maximizes a modularity function, which measures the relative density of edges inside communities with respect to edges outside them. It matches the goal of optimizing the modularity of the whole binary. That is, each module explored holds high modularity. (2) It scales to very large graphs within limited computing and time resources, critical for binaries having many functions.
\looseness=-1

For the second condition, a module generated by Louvain may contain a large number of functions, exceeding the token limit of LLM. We thus introduce the hierarchical strategy. If a module violates a pre-defined constraint to LLM, we recursively apply Louvain algorithm to further decompose it into sub-modules until the constraint is met. The strategy preserves the module's original program behavior through hierarchical relationships (i.e., sub-program behaviors) and satisfies the token limits. \looseness=-1

Specifically, we define a \textit{function number constraint}, each module having at most $n$ functions where $n$ is dynamically determined through the size $S$ of the full context window of LLM. The total input token number $T_f$ filled by the module with $n$ functions must satisfy: \looseness=-1

\begin{equation}
    T_f \leq S - R_m - P_m
\end{equation}

\noindent where $P_m$ is the token number of the pre-defined content in the prompt template for module summarization, and $R_m$ is the reserved token space used to generate the module summary (set to 0.5K). In addition, our statistical analysis reveals that pseudo functions consume approximately 2-4 times the tokens of source functions (with an average of 556 tokens~\cite{keenhash:2025}). Thus, $n$ can be determined around 50 -- 100, costing 111K -- 222K tokens on average, for LLM inferring (e.g., in a context window of 272K~\cite{openai}) in one conversation. This scale is also sufficient to capture a (sub-)program behavior.

\begin{algorithm}[t]
\normalem
\caption{The HMD Algorithm}
\label{alg:hmd}

\SetKwFunction{FDecompose}{Decompose}
\SetKwProg{Fn}{Function}{:}{}

\KwIn{$BG$ -- Binary graph, $Threshold$ -- \textit{function number constraint};}
\KwOut{$\mathcal{M}$ -- Generated modules, $\mathcal{H}$ -- Hierarchical tree structure;}

\BlankLine

$\mathcal{M}, \mathcal{H} \gets$ \textsf{Decompose}($BG$, $Threshold$, $1$);

\Return{$\mathcal{M}$, $\mathcal{H}$};

\Fn{\FDecompose{$G$, $Threshold$, $Level$}}{
    $\mathcal{O}, \mathcal{Q} \gets \emptyset, \emptyset$;

    $\mathcal{R} \gets $\textsf{Louvain}($G$); \tcp*[r]{Apply Louvain algorithm to generate modules}

    \ForEach{$m_i$ \textbf{\text{in}} $\mathcal{R}$}{
        $\mathcal{Q} \gets $ $\mathcal{Q} \cup \textcolor{white}{}$ \textsf{Edge}($G$, $m_i$, $Level$);  \tcp*[r]{Record hierarchy information (with edge), including levels}

        \tcp*[l]{\textsf{Token}($m_i$) computes the total input token number for $m_i$ and $|m_i|$ represents the number of functions in $m_i$}
        \eIf{\textsf{Token}($m_i$) $> Threshold$ \textbf{\text{and}} $|m_i| > 1$}{  
            $\mathcal{O}_{sub}, \mathcal{Q}_{sub} \gets \text{\textcolor{white}{}}$ \textsf{Decompose}($m_i$, $Threshold$, $Level + 1$); \tcp*[r]{Recursive decomposition for unsatisfied modules}
        
            $\mathcal{O}, \mathcal{Q} \gets \mathcal{O} \cup \mathcal{O}_{sub}, \mathcal{Q} \cup \mathcal{Q}_{sub}$;  \tcp*[l]{Add leaf-node modules and hierarchy information}
        }{
            $\mathcal{O} \gets \mathcal{O} \cup \{m_i\}$;  \tcp*[r]{Add leaf-node module}
        }
    }
    \Return{$\mathcal{O}$, $\mathcal{Q}$};
}
\end{algorithm}

As shown in Algorithm~\ref{alg:hmd} for \textit{Hierarchical Modular Decomposition} (HMD), we apply Louvain to binary graph to maximize the modularity function (lines 1 and 5)~\cite{dugue2015directed}:

\begin{equation}
Q = \frac{1}{m} \sum_{i=1}^{N} \sum_{j=1}^{N} \left[ A_{ij} - \frac{k_i^{in} k_j^{out}}{m} \right] \delta(C_i, C_j)
\end{equation}

\noindent where $A_{ij}$ is the correlation between vertex $i$ and $j$ in the binary graph; $m$ stands for the number of edges; $N$ is the number of functions; $k_i^{in}$/$k_j^{out}$ stands for the in/out-degree of $i$/$j$; and, $C_{i\text{/}j}$ is the module to which $i$/$j$ belongs and $\delta(u, v)$ is defined as 1 if $u=v$, and 0 otherwise.
The weights of all edges are considered equivalent, reflecting the equal significance of the four types of graphs.

If a generated top-level (i.e., level 1) module does not satisfy the \textit{function number constraint} and the number of functions $> 1$ (line 8), then we recursively run Louvain, i.e., function \textsf{Decompose}, on the module until all decomposed sub-modules (i.e., level 2--$i$) satisfy the constraint (lines 6--14).
The process forms a hierarchical tree structure, and such hierarchy information $\mathcal{H}$ is recorded accordingly (lines 7, 9, and 10). Eventually, we take all the fine-granularity leaf nodes, only with their functions, as the final generated modules $m_i \in \mathcal{M}$. These modules are then passed to \textit{Modular Optimization and Summarization}. \looseness=-1

\subsubsection{Modular Optimization and Summarization}\label{sec:modularoptimizationandsummarization}

In the second step, \binxsight{} summarizes the generated modules.
As noted, a large binary can yield numerous modules (e.g., 6,582 for Z3), many of which are non-essential.
We thus propose a MapReduce approach for efficient module summarization.
\looseness=-1

\smallskip
\noindent \textbf{Feature-oriented Module Ranking and Filtering.} Only some modules contribute to a binary's core behaviors. For example, gh0st RAT malware~\cite{gh0st} implements a memory manager to manage buffer allocation, which does not directly reflect the core remote control behaviors. In this case, summarizing modules related to memory management is of low priority. Thus, discarding them not only prevents the introduction of irrelevant information but also reduces the resource costs. Based on this insight, we design a strategy to rank and filter high-value modules.
\looseness=-1

$\bullet$ \textbf{\textit{Ranking:}} We use two heuristic features to describe the importance of modules for ranking: (1) the average lines of code of functions (LoC), and (2) the proportion of user-developed functions (PoU), which are defined as follows:\looseness=-1

\begin{equation}
\begin{split}
\text{LoC}(m_i) &= \frac{1}{|m_i|} \sum_{f \in m_i} \text{LoC}(f),\\
\text{PoU}(m_i) &= \frac{|\{ f \in m_i \mid f \text{ is user-developed} \}|}{|m_i|}
\end{split}
\end{equation}

\noindent where $f$ denotes the function in module $m_i \in \mathcal{M}$.
Following previous work~\cite{keenhash:2025, yang2022modx}, a higher LoC likely handles more complex logic and presents some core behaviors, making it crucial within the program. Furthermore, a greater PoU tends to represent the semantics of user-developed code.
We rank modules $m_i$ using a lexicographic ordering, prioritizing $\text{LoC} \mid T$ first, and then PoU:

\begin{equation}
\begin{split}
P(m_i) \prec P(m_j)
\iff
\Big(
 \text{LoC}(m_i) \mid T <
 \text{LoC}(m_j) \mid T 
\Big)
\\
\ \vee\
\Big(
\text{LoC}(m_i) \mid T =
\text{LoC}(m_j) \mid T
\wedge
\text{PoU}(m_i) < \text{PoU}(m_j)
\Big)
\end{split}
\end{equation}

\noindent Where, $P(m_i)$ denotes the priority of $m_i$. The average code size, compared to the proportion of user-developed code, more directly represents the behavioral semantics. Thereby, we place LoC higher than PoU.
In addition, LoC is divided by $T$ to prioritize the semantics of user-developed code within the corresponding intervals ($T$ is set to 50 by default).
\looseness=-1

$\bullet$ \textbf{\textit{Filtering:}} By limiting an appropriate input token resource regarding the program summarization, we can filter the high-value feature-optimized modules, thus achieving an effective, fast, and economical process.
We customize the LLM's context window size (CWS) $S_p$ in \textit{Program Prompting} (Sec. \ref{sec:programprompting}) as the filtering criterion. 
Since modules are approximately isolated (Sec.~\ref{sec:motivation}), we adapt the \textit{MapReduce strategy}~\cite{dean2008mapreduce} to perform filtering and per-module summarization in parallel.
\looseness=-1

Specifically, \binxsight{} leverages \textit{Leaf-node Module Summarization} (see later) to summarize ranked modules from $\mathcal{M}$ in a batch manner (\textit{Map Phase}). Meanwhile, it continuously accumulates the tokens of the generated module summaries. When the cumulative token number exceeds a threshold after adding the tokens of the ranked $i_{th}$ module, we discard it and all the remaining unprocessed batches. For the current batch, we sequentially attempt to retain summaries starting from the $i_{th} + 1$ module until all remaining module summaries have been scanned. The filtered modules are denoted as $\mathcal{M}'$.
The threshold is defined as $S_p - R_p - P_p$ where $P_p$ is the token number of the pre-defined content in the prompt template for the final program summarization and $R_p$ is the reserved token space, with 0.5K, used to generate the final program summary. \looseness=-1

\smallskip
\noindent \textbf{Leaf-node Module Summarization.} 
Summarizing functions first, as in existing function-level BCS~\cite{ye2023cp, jin2023binary, zhu2025misum}, prevents the LLM from observing complete calling relationships, hindering its inference of both function and module semantics.
For example, a WannaRen ransomware decryption tool~\cite{ransomdecrypt} contains a decryption function, yet function-level summarization misreads it as encryption and propagates this error to the module and final summaries.
\looseness=-1

To better summarize a module, we prompt all the functions from the module to LLM once (i.e., one-time conversational modular prompting), along with control flow and data reference information. Thus, LLM can observe all function information and its dependencies for better inferring.
For instance, the decryption function in the previous ransomware decryption tool can be correctly inferred to be the decryption algorithm for the encrypted files based on its context.
Moreover, to avoid introducing information unrelated to behavior and facilitate the later Reduce phase, we carefully design our prompt template, including system prompt and user prompt:

\begin{Verbatim}[
    frame=single,
    baselinestretch=1,
    fontsize=\footnotesize,
    xleftmargin=0.ex,
    breaklines=true,
    commandchars=\\\{\}
]
\PYG{esc}{\underline{\# System Prompt:}}
\PYG{esc}{You are a binary security analysis expert who can assist users in analyzing the functionalities of binary programs.}
\PYG{esc}{I will provide the pseudo functions of a binary program module decompiled through Ghidra, along with the control dependency information in the module.}
\PYG{esc}{Your task is to summarize the overall behavior of this binary program module based on all the information provided.}
\PYG{esc}{\textbf{Requirements:}}
\PYG{esc}{\textbf{1. Provide a concise and accurate summary of the module's overall behavior with no explanation;}}
\PYG{esc}{\textbf{2. Do not delve into the implementation details of provided pseudo functions and the dependency information.}}
\PYG{esc}{\underline{\# User Prompt:}}
\PYG{esc}{The pseudo functions along with the control dependency information in the module are provided as follows:}
\PYG{esc}{\textcolor{red}{[\textbf{Pseudo Functions}]}}, \PYG{esc}{\textcolor{red}{[\textbf{Control Flow}]}}
\end{Verbatim}

The system prompt provides a brief overview of the task and the inferring requirements. The user prompt gives the specific module information including pseudo functions and corresponding control flow. Functions are provided in a format of <\textit{function name}, \textit{pseudo function content}>. The dependency is provided in JSON graph. We also comment function names of the callees at the indirect calls within the pseudo function.

Additionally, to help LLM better comprehend the semantics of functions, we perform function name prediction (FNP) to provide hints. We leverage function clone search~\cite{wang2024clap, wang2022jtrans} (0.75 as the threshold) in a binary-to-source manner to predict function names. Specifically, we choose the state-of-the-art KEENHash function embedding model~\cite{keenhash:2025} to represent both pseudo and source functions and build a repository (following \cite{keenhash:2025}) having only source functions for prediction. We further discuss the design choice and the accuracy of our FNP in Sec.~\ref{sec:discussion}.
\looseness=-1

\smallskip
\noindent \textbf{Reconstruction.} As modules are ranked and filtered in the previous process, the hierarchy information of modules below level 1 (i.e., level 2-$i$. See Fig. \ref{fig:modular-sum}) can be fragmented. Directly providing the generated module summaries to LLM for final program summarization can make it difficult to understand which level 2-$i$ sub-program behaviors contribute to a specific level 1 program behavior.
Therefore, we leverage the hierarchy information extracted from \textit{Hierarchical Modular Decomposition} (Sec. \ref{sec:hierarchicalmodulardecomposition}) to reconstruct the module summaries of $\mathcal{M}'$, letting LLM better understand a specific level 1 program behavior.
For each level 1 node or module having level 2-$i$ leaf-node modules, \binxsight{} nests and reorganizes their module summaries according to the hierarchical tree structure $\mathcal{H}$ if they are not filtered out, e.g., [[$sum_2^1$, $sum_2^2$, [$sum_3^1$, $sum_3^2$, $sum_3^3$]]] as the level 1 program behavior (i.e., level 1 module summary) where $sum_i^j$ represents level $i$'s $j_{th}$ module summary. Eventually, these reconstructed level 1 module summaries $\mathcal{M}'_{sum}$ are passed to \textit{Program Prompting} for the final program summarization.
\looseness=-1

\subsubsection{Program Prompting}\label{sec:programprompting}

In the final step, \binxsight{} summarizes the entire binary. As module summaries provide approximately isolated abstractions of program behaviors, it is suitable to extract the functionality from them directly. Specifically, we merge all module summaries $\mathcal{M}'_{sum}$ and send them to LLM for generating the final \textit{program summary} in a one-time prompting manner (\textit{Reduce Phase})~\cite{summarizetext}. We carefully design the prompt template as follows:

\begin{Verbatim}[
    frame=single,
    baselinestretch=1,
    fontsize=\footnotesize,
    xleftmargin=0.ex,
    breaklines=true,
    commandchars=\\\{\}
]
\PYG{esc}{\underline{\# System Prompt:}}
\PYG{esc}{I will provide a behavioral summary of each module of a binary program. Your task is to summarize the functionality of this binary program based on all its modules.}
\PYG{esc}{\textbf{Requirements:}}
\PYG{esc}{\textbf{1. Provide a concise and accurate summary of the binary program's overall functionality with no explanation;}}
\PYG{esc}{\textbf{2. The summaries enclosed in '[]' correspond to a specific program behavior (i.e., one module identity) of the binary.}}
\PYG{esc}{\underline{\# User Prompt:}}
\PYG{esc}{Here are all the modules' behaviors regarding the program:}
\PYG{esc}{\textcolor{red}{[\textbf{Level 1 Module Summaries}]}}
\end{Verbatim}

The system prompt provides a brief overview of the task and requires LLM to generate a concise and accurate summary. The user prompt gives the specific reconstructed level 1 module summaries, in a format of <\textit{module identity}, \textit{level 1 module summary content}>.

%% file: secs/evaluation.tex
\section{Evaluation}\label{sec:evaluation}

We implement \binxsight{} on Ghidra~\cite{ghidra} and a widely used value-set analysis framework~\cite{binabsinspector}. The binary graph construction is built using NetworkX~\cite{networkx}, KEENHash~\cite{keenhash:2025}, FAISS~\cite{douze2024faiss}, and BinaryAI~\cite{jiang2024binaryai}. The hierarchical summarization is developed with NetworkX and openai-python~\cite{openaipython}, leveraging KEENHash and FAISS for function name prediction.
In this section, we evaluate \binxsight{} by answering the following research questions:

\noindent $\bullet$ \textbf{RQ1:} How effective is \binxsight{} for program-level BCS?

\noindent $\bullet$ \textbf{RQ2:} How efficient is \binxsight{} for program-level BCS?  

\noindent $\bullet$ \textbf{RQ3:} How do users perceive the usability of \binxsight{}?

\noindent $\bullet$ \textbf{RQ4:} How does each component contribute to \binxsight{}?

\noindent Additionally, we perform a case study on real-world malware analysis in Sec. \ref{sec:caseanalysis}.

\subsection{Dataset}\label{sec:dataset}

\noindent \textbf{Test Dataset.} To thoroughly evaluate \binxsight{}, we collect 92 popular or notorious (i.e., malware) C/C++ projects (Table~\ref{tab:dataset}). They cover diverse fields, such as network communication, remote control malware, gaming, DDoS virus, and so forth, across various platforms, including Windows, Linux, and macOS.
For each project, we directly use the latest released version of the artifact. If there are multiple architectures for the artifact, we select one of them randomly. As for the projects with no artifact, we compile them in the default optimization level O2 with a randomly selected architecture from x86, x64, ARM, and MIPS.
\looseness=-1

\begin{table}[t]
    \centering
    \caption{92 projects and 110 binaries in our dataset.}
    \label{tab:dataset}
    \setlength{\tabcolsep}{3pt}
    \resizebox{\textwidth}{!}{
    \begin{tabular}{cc||cc||cc||cc}
    \toprule
    \textbf{Project} & \textbf{\# Binaries} & \textbf{Project} & \textbf{\# Binaries} & \textbf{Project} & \textbf{\# Binaries} & \textbf{Project} & \textbf{\# Binaries} \\
    \midrule

    adns & 3 & aspell & 1 & awk & 1 & bash & 1 \\
    bison & 1 & cflow & 1 & controlblock & 1 & cpio & 1 \\
    cppi & 1 & cryptolocker & 1 & ctags & 1 & curl & 1 \\
    datamash & 1 & db-bench & 1 & direvent & 1 & ecal-rec & 1 \\
    enscript & 1 & st-device-sdk-c & 1 & ffmpeg & 1 & freetype & 1 \\
    gcal & 1 & gdbm & 3 & gh0st & 3 & grep & 1 \\
    gzip & 1 & kvrocks & 1 & less & 1 & a2ps & 1 \\
    audioflux & 1 & glpk & 1 & gmp & 1 & gnudos & 1 \\
    gsasl & 1 & gsl & 1 & guile & 1 & idn & 1 \\
    libjpeg & 1 & lightning & 1 & magickwand & 1 & libmicrohttpd & 1 \\
    mpfr & 1 & nettle & 1 & osip & 2 & libpng & 1 \\
    readline & 1 & sqlite & 1 & libtasn1 & 1 & libunistring & 1 \\
    libxml2 & 1 & linux-ddos & 1 & linux-encoder & 2 & linux-wirenet & 1 \\
    llama2.c & 2 & macchanger & 1 & make & 1 & masscan & 1 \\
    minhook & 1 & mirai & 2 & nano & 1 & nano-node & 1 \\
    objcopy & 1 & objdump & 1 & openssl & 1 & patch & 1 \\
    poke & 1 & ransom-wannaren-decryption & 1 & ransomware-wannacry & 1 & ransomware-xdata & 1 \\
    rbot & 1 & readelf & 1 & replay-sorcery & 1 & sed & 1 \\
    ssdeep & 1 & openssh & 3 & sshd-malware & 5 & tar & 1 \\
    tcpdump & 1 & tic80 & 1 & time & 1 & tmux & 1 \\
    tortoisegitmerge & 1 & tortoisegitproc & 1 & ujson & 1 & units & 1 \\
    unzip & 1 & wdiff & 1 & wget2 & 1 & whisper & 3 \\
    winsparkle & 1 & xmrig & 1 & xorriso & 1 & z3 & 1 \\

    \bottomrule
    \end{tabular}
    }
\end{table}

\begin{figure}[t]
\centering
\includegraphics[width=0.6\textwidth]{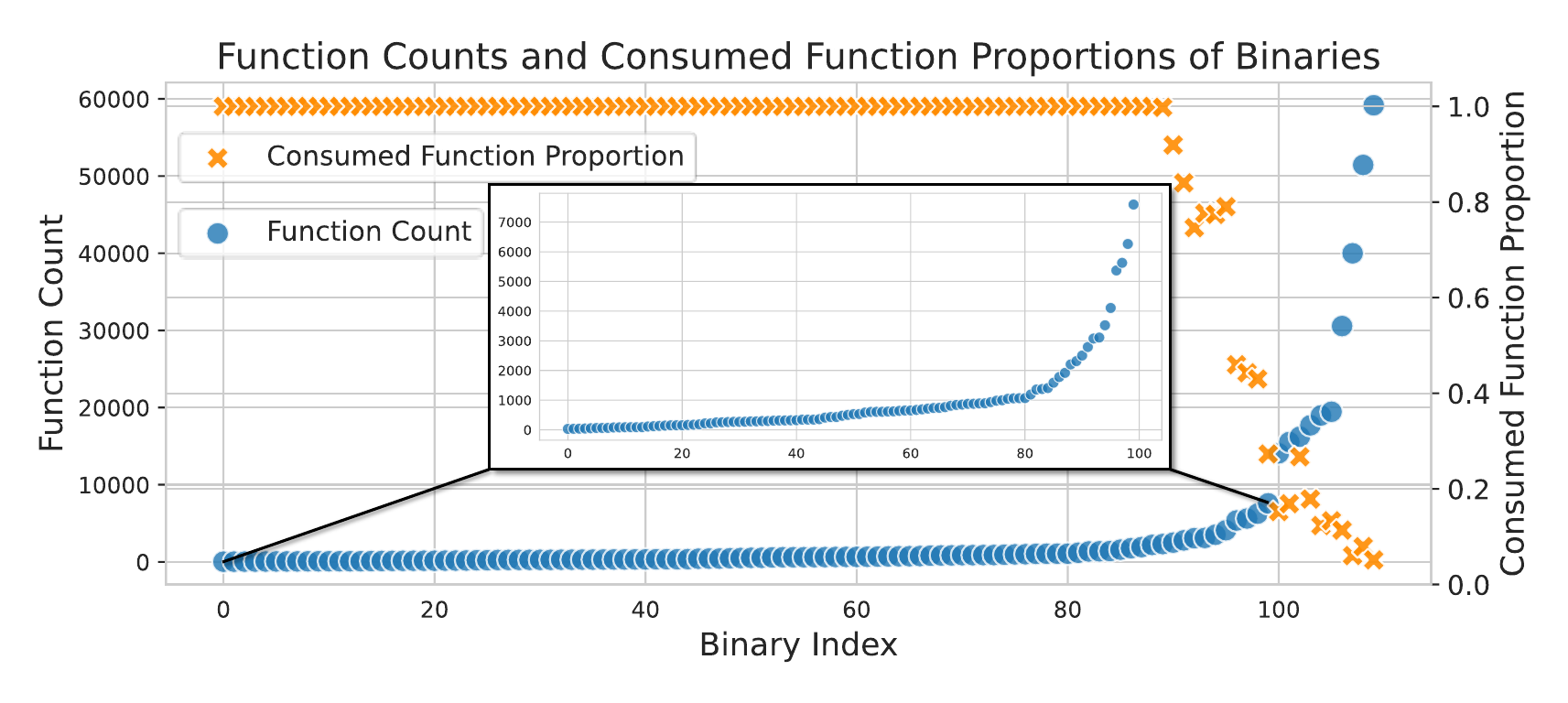}
\caption{The function counts and the consumed function proportions of the binaries in the test dataset.}
\label{fig:function-size}
\end{figure}

In total, we collect 110 binaries (91 benign and 19 malicious samples). All the binaries are stripped. Moreover, we plot the function counts of these binaries in Fig. \ref{fig:function-size}. The variation in the function counts (31 to 59,173 functions, on average of 3,429) allows for the evaluation of generalizability for \binxsight{} challenging different sizes of binaries. \looseness=-1

\smallskip
\noindent \textbf{Ground Truth for Test Dataset.} Constructing a high-quality ground-truth summary to each binary program is key to effectively evaluating program-level BCS. However, unlike function-level BCS having meaningful comments or names as ground truth~\cite{zhu2025misum, ye2023cp, hu2024degpt, jin2023binary}, the projects of the binaries lack directly usable ones. Additionally, binary programs are composed of numerous functions, and thus, providing their comprehensive ground truth is non-trivial. To this end, we employ a three-phase method combining LLM with manual efforts to generate high-quality ground truth, following the guideline in Sec. \ref{sec:definitionofbps}.

Specifically, we first collect binaries' \code{README.md} documents and all comments in source code. The \code{README.md} typically contains a project overview and key features, while comments include functional descriptions for functions or compile units (i.e., source file). Thus, both of them together serve as the foundation for subsequent ground truth generation. In addition, for binaries without \code{README.md} and source code (i.e., in-the-wild malware in our dataset), we upload them to VirusTotal~\cite{virustotal}, the world's biggest threat intelligence company, and obtain their relevant threat reports (analyzed by human experts from security vendors) as substitutes. Next, for each binary, we send its \code{README.md} and comments (or threat reports) to an LLM (with \textit{refine} strategy~\cite{summarizetext}) and request it to summarize the functionality of the binary in a concise and accurate single paragraph. Finally, we have three human experts, each with at least seven years of experience in reverse engineering and software development, cross-check the ground-truth summaries of the binaries. In particular, one expert supplements and revises the LLM-generated ground truth, while two other experts perform the confirmation. This process continues until all ground-truth summaries are deemed to comprehensively cover the functionality of the binaries. In total, we spend \textit{\textit{336 person-hours}} to ensure the quality of the ground truth. Furthermore, in this study, we leverage gpt-5.4~\cite{openai} to generate the initial ground-truth summaries.

\subsection{Experiment Setup}\label{sec:experimentsetup}

\noindent \textbf{\binxsight{}.} We choose gpt-5.4~\cite{openai}, with a context window size of 272K token limits, as the base LLM.
We explore the effectiveness of \binxsight{} (and baselines) supported by other LLMs (e.g., claude-opus-4-6~\cite{claude-opus-4-6}) in Sec.~\ref{sec:ablationstudyandcomponentanalysis}.
The temperature is set to 0 to stabilize results and ensure reproducibility.
The batch size is set to 40 for parallel module summarization.
We set the filtering criterion $S_p$ (see Sec. \ref{sec:modularoptimizationandsummarization}) with 16K token limits, as it optimally balances the performance, time consumption, and economic costs of \binxsight{}. The exploration of the different sizes of $S_p$ is shown in Sec. \ref{sec:ablationstudyandcomponentanalysis}. 
For function name prediction (FNP) in Sec.~\ref{sec:modularoptimizationandsummarization}, we construct a source function repository for clone search.
In particular, we follow previous studies~\cite{tang2022towards, wu2023ossfp, keenhash:2025} to collect a large number of diverse open-source C/C++ projects by crawling from Github~\cite{github} and GNU/Linux community~\cite{gnulinuxcommunity}. In total, 11,126 projects are obtained, containing 57M unique C/C++ source functions as the repository (functions appearing in our test dataset are excluded based on the hashes of their normalized source code).

\smallskip
\noindent \textbf{Baseline.} To the best of our knowledge, all the previous works~\cite{ye2023cp, jin2023binary, lu2024malsight, zhu2025misum, symgen:2025, jin2022symlm} focus only on function-level BCS. Thus, we adjust them to adapt to the program level. In total, we have 5 baselines:

\noindent $\bullet$ \textit{func-sum:} It first summarizes pseudo functions, sorted by \textit{rva}, and then leverages function summaries to generate the final program summary. It also supports MapReduce and filtering. We implement it on gpt-5.4 (the same applies to other baselines by default). 

\noindent $\bullet$ \textit{func-sum-MiSum:} We incorporate the state-of-the-art function-level BCS method, MiSum~\cite{zhu2025misum}. In this setting, we replace the gpt-5.4-based function summarization with MiSum in func-sum.

\noindent $\bullet$ \textit{func-code:} It sends pseudo functions to LLM directly in one conversation for program summairzation. It is a variant to \binxsight{} without modular consideration.

\noindent $\bullet$ \textit{sym-sum:} sym-sum first predicts function names to all functions. Then, it sends all the predicted names to LLM in one conversation for program summarization. We utilize the state-of-the-art LLM-powered method SymGen~\cite{symgen:2025} to predict function names.

\noindent $\bullet$ \textit{Malsight:} It~\cite{lu2024malsight} extracts the call graph and continuously summarizes functions with LLM from leaf nodes to the root while carrying previously summarized information. The summary of the root can be approximated as the program summary. If there are multiple roots, we perform another round of information extraction on them.

\noindent For func-sum (func-sum-MiSum), func-code, and sym-sum, we set their batch sizes as 40 (128), 40, and 128.

\smallskip
\noindent \textbf{Evaluation Metric.} We adopt 4 metrics to ensure a comprehensive evaluation, including 1 embedding-based metric (SIMILARITY) of text-embedding-3-large~\cite{openai} and 3 widely-used traditional metrics of BLEU~\cite{papineni2002bleu}, ROUGE-L~\cite{lin2004rouge}, and METEOR~\cite{banerjee2005meteor}.
The embedding metric evaluates the semantic similarity of summaries~\cite{jin2024simllm} by encoding text into cosine similarity space (range: $[-1, 1]$).
To select the embedding model, we compare text-embedding-3-large with two other recent embedding models: the general-purpose voyage-3-large~\cite{voyage3large} and the state-of-the-art SimLLM~\cite{jin2024simllm} designed for code summarization.
Following the experimental design of \cite{jin2024simllm}, we measure how well the similarity scores produced by each model between ML-generated code summaries and the ground truth align with human expert ratings.
As shown in Table \ref{tab:metric}, text-embedding-3-large achieves the highest Spearman’s rank correlation coefficients on both code summary datasets~\cite{haque2022semantic, shen2021evaluating} (e.g., 0.867 on Haque~\cite{haque2022semantic}), indicating the strongest alignment with human ratings.
Thus, we adopt it as the embedding metric for evaluation.
\looseness=-1

\begin{table}[t]
    \centering
    \caption{Spearman’s rank correlation coefficient of evaluation metrics on Haque~\cite{haque2022semantic} and Shen~\cite{shen2021evaluating} datasets. OpenAI and Voyage represent text-embedding-3-large and voyage-3-large, respectively.}
    \label{tab:metric}
    \scalebox{0.75}{
    \begin{tabular}{c|c|c|c|c|c|c}
    \toprule
    \multirow{2}{*}{\textbf{Dataset}} &  \multicolumn{6}{c}{\textbf{Metric}}  \\
    \cmidrule{2-7}
    & OpenAI~\cite{openai} & Voyage~\cite{voyage3large} & SimLLM & BLEU & ROUGE-L & METEOR \\
    \midrule
    \multicolumn{1}{c|}{Haque~\cite{haque2022semantic}} & \textbf{0.867} & 0.828 & 0.837 & 0.719  & 0.800 & 0.751 \\
    \multicolumn{1}{c|}{Shen~\cite{shen2021evaluating}} & \textbf{0.765} & 0.747 & \textbf{0.765} & -0.272 & 0.758 & 0.740 \\
    \bottomrule
    \end{tabular}
    }
\end{table}

\smallskip
\noindent \textbf{Experiment Environment.} All the experiments are run on a Linux server running Ubuntu 20.04 with AMD EPYC 7K62 48-Core Processor, 1TB RAM, and 8 Nvidia A100 GPUs.

\subsection{RQ1: Performance of \binxsight{}}\label{sec:performance}

\noindent \textbf{Motivation.} In the first RQ, we evaluate the summarization performance of \binxsight{}.

\smallskip
\noindent \textbf{Approach.} We perform program-level BCS on the test dataset, comparing \binxsight{} with the 5 baseline methods.

\smallskip
\noindent \textbf{Result.} Our results (see Table \ref{tab:performance}) show that \binxsight{} achieves the best overall performance with 0.6137 SIMILARITY, 4.6204 BLEU, 0.1897 ROUGE-L, and 0.2548 METEOR, significantly outperforming the 5 baselines. This indicates that \binxsight{} summarizes binaries more effectively, and its high SIMILARITY score also reflects its ability to capture the functionality of binaries faithfully.

\begin{table}[t]
    \centering
    \caption{The performance of \binxsight{}, compared with 5 baselines, on SIMILARITY, BLEU, ROUGE-L, and METEOR metrics.}
    \label{tab:performance}
    \scalebox{0.75}{
    \begin{tabular}{c|c|c|c|c}
    \toprule
    \textbf{Method} & \textbf{SIMILARITY} &\textbf{BLEU} & \textbf{ROUGE-L} & \textbf{METEOR} \\
    \midrule

    func-sum       & 0.4688 & 2.9170   & 0.1421   & 0.2165 \\
    func-sum-MiSum & 0.4175 & 2.1135   & 0.1274   & 0.1936 \\
    func-code    & 0.4982 & 3.2273   & 0.1457   & 0.2196 \\
    sym-sum      & 0.4670 & 2.8945   & 0.1402   & 0.2112 \\
    Malsight     & 0.4814 & 3.2461   & 0.1508   & 0.2175 \\
    \binxsight{} & \textbf{0.6137 ($\uparrow \text{23.18\% - 46.99\%}$)} & \textbf{4.6204}   & \textbf{0.1897}   & \textbf{0.2548} \\

    \bottomrule
    \end{tabular}
    }
\end{table}

\noindent $\bullet$ \textit{Comparison with func-sum:} \binxsight{} outperforms func-sum by 30.91\%, 58.40\%, 33.50\%, and 17.69\% in SIMILARITY, BLEU, ROUGE-L, and METEOR. Regarding SIMILARITY, the significant gap indicates that \binxsight{} possesses a more comprehensive semantic summarization, easier to understand the functionality in reverse engineering tasks, e.g., malware analysis~\cite{keenhash:2025}. func-sum is based on function summary, but function-level granularity does not allow for the observation of contextual information, resulting in the loss of useful semantics. Consequently, the generated program summaries are less effective. While \binxsight{} captures essential functionality through modules, minimizing information loss to provide more comprehensive summaries of binaries. For example, \binxsight{} can understand Mirai's DDoS attack chain (Fig. \ref{fig:attack-func-sim}), but func-sum only recognizes behaviors involving network requests.

\noindent $\bullet$ \textit{Comparison with func-sum-MiSum:} \binxsight{} outperforms func-sum-MiSum by 46.99\%, 118.61\%, 48.90\%, and 31.61\% in the four metrics. As MiSum relies on a limited training set for function-level BCS, it cannot generalize to arbitrary binaries, and its performance is even worse than func-sum in program-level BCS.

\noindent $\bullet$ \textit{Comparison with func-code:} \binxsight{} outperforms func-code by 23.18\%, 43.17\%, 30.20\%, and 16.03\% in the four metrics. func-code sends pseudo functions directly to LLM, minimizing information loss in function summarization, compared with func-sum. However, the lack of modular decomposition for grouping relevant functions and the constraint of the context window size make it difficult for func-code to summarize comprehensive functionality. In fact, the summary results of \binxsight{} for each binary are consistently higher than those of func-code in SIMILARITY, even for \binxsight{}-w/o-FNP (remove FNP) in Sec. \ref{sec:ablationstudyandcomponentanalysis}.
\looseness=-1

\noindent $\bullet$ \textit{Comparison with sym-sum:} \binxsight{} outperforms sym-sum by 31.41\%, 59.63\%, 35.31\%, and 20.64\%. Although sym-sum can observe a sufficient number of function names, the recovered results may be abstract (e.g., \code{run}) or misleading, making the summarization less effective.

\noindent $\bullet$ \textit{Comparison with Malsight:} \binxsight{} also outperforms Malsight by 27.48\%, 42.34\%, 25.80\%, and 17.15\%. Such a result indicates that \binxsight{}'s binary graph construction and hierarchical summarization can more comprehensively extract functionality from binaries.

\subsection{RQ2: Efficiency of \binxsight{}}\label{sec:efficiency}

\noindent \textbf{Motivation.} One of the challenges for program-level BCS is to summarize binaries within a reasonable time. In this RQ, we evaluate the efficiency of \binxsight{}.

\smallskip
\noindent \textbf{Approach.} We summarize the 110 binaries one by one while recording the time spent.
We exclude the time spent on offline binary processing (e.g., decompilation), measuring only the summarization process here.

\smallskip
\noindent \textbf{Result.} The results are shown in Table \ref{tab:time}. \binxsight{} takes only 44.6 minutes to summarize all binaries (i.e., 24.3 seconds on average). It is significantly faster than func-sum's 113.4 minutes, sym-sum's 174.2 minutes, and Malsight's 15.2 days, comparable to func-sum-MiSum's 51.8 minutes, but slower than func-code's 15.3 minutes.

\begin{table}[t]
    \centering
    \caption{The time complexity and consumption of \binxsight{} and baselines for summarizing the test dataset.}
    \label{tab:time}
    \scalebox{0.75}{
    \begin{tabular}{c|c|c|c|c}
    \toprule
    \textbf{Method} & \textbf{Time Complexity} &\textbf{Min} & \textbf{Max} & \textbf{Total} \\
    \midrule

    func-sum        & $O(1)$ & 6.2 sec  & 4.6 min  & 113.4 min \\
    func-sum-MiSum  & $O(1)$ & 5.3 sec  & 3.2 min  & 51.8 min \\
    func-code    & $O(1)$ & 4.1 sec  & 22.9 sec & 15.3 min \\
    sym-sum      & $O(1)$ & 5.9 sec  & 25.1 min & 174.2 min \\
    Malsight     & $O(n)$ & 87.4 sec & 2.6 days & 15.2 days \\
    \binxsight{} & $O(1)$ & 5.7 sec  & 3.8 min  & 44.6 min \\

    \bottomrule
    \end{tabular}
    }
\end{table}

Compared with func-code, \binxsight{} and the other 3 methods (i.e., func-sum, func-sum-MiSum, and sym-sum) rely on two-phase summarization rather than a single one, which results in a slower procedure. 
\binxsight{} is faster than func-sum and sym-sum since it summarizes at the module granularity and limits the context window size $S_p$ for filtering, resulting in a total number of modules to summarize being much less than that of functions.
For instance, \binxsight{} requests LLM 133 times for the modules of Z3 (59,173 functions), while func-sum and sym-sum have 2,059 and 43,114 times.
As a result, \binxsight{} costs a maximum of 3.8 minutes for Z3, but func-sum and sym-sum take 4.6 and 25.1. 
Moreover, \binxsight{} is comparable to func-sum-MiSum. This is because the latter uses a traditional DL model, which can generate function summaries more quickly.
In addition, we also count the consumed function proportions to binaries by \binxsight{} in Fig. \ref{fig:function-size}. It employs the filtering strategy to preserve high-value modules and can discard the vast majority of low-value ones, especially when summarizing large binaries and thereby ensuring efficiency.

As for Malsight, it takes at least 87.4 seconds to summarize a small binary, while summarizing a binary with more functions can even exceed 24 hours (e.g., Z3's 2.6 days). Its tree-structured serial summarization nature renders it inefficient when handling large binaries, with a time complexity of $O(n)$ where $n$ is the function counts. However, for the other methods, their parallelizable nature and filtering strategy enable them to perform fast summarization with a time complexity of $O(1)$. \looseness=-1

\subsection{RQ3: User Study}\label{sec:userstudy}

\noindent \textbf{Motivation.} \binxsight{} pays attention to the practical effects of assisting the reverse engineers in understanding binaries. Here, we evaluate the extent to which it helps them using a question-based user study. The IRB of our affiliation has approved our study.

\smallskip
\noindent \textbf{Study Preparation.} We invited \textbf{\textit{27}} participants to explore the practical value of \binxsight{} for reverse engineers in understanding unknown binaries.
The study participants include 15 seasoned reverse engineers from leading security vendors, each with a minimum of 7 years of hands-on experience in reverse engineering, alongside 12 PhD candidates specializing in security research, all of whom possess at least 3 years of practical experience.
Furthermore, we randomly choose 10 tuples of <stripped binary, first program summary, second program summary, VirusTotal URL, Project URL/Threat Reports> for each participant and query them for scoring, focusing on the following questions (Q1-Q5).
In which case, two summaries come from \binxsight{}, and the baseline achieving the highest SIMILARITY on that binary among the 5 baselines, and they are randomly assigned to either the first or the second to eliminate bias.
We provide a tip: "\textit{Please rate on a scale of 0 to 10, with higher scores indicating better quality.}” to guide the scoring of each question for the two summaries.
The procedure employs a method of first conducting independent reverse engineering, followed by scoring.
In total, we collect 270 $\times$ 2 scores for each question.
\looseness=-1

\noindent $\bullet$ \textit{Q1 (Meaningfulness and Helpfulness):} Do you think the summary is meaningful and helpful for understanding the binary?

\noindent $\bullet$ \textit{Q2 (Minimal Superfluous Description):} Do you think the summary has no or little superfluous descriptions that have nothing to do with the functionality of the binary itself?

\noindent $\bullet$ \textit{Q3 (Readability):} Do you think the summary is clear, concise, and easy to understand the binary?

\noindent $\bullet$ \textit{Q4 (Functionality Understanding):} Do you think the summary gives a good understanding of the functionality of the binary?

\noindent $\bullet$ \textit{Q5 (Completeness):} Do you think the summary is complete (e.g., is there core functional information missing)?

\begin{figure}[t]
\centering
\includegraphics[width=0.6\textwidth]{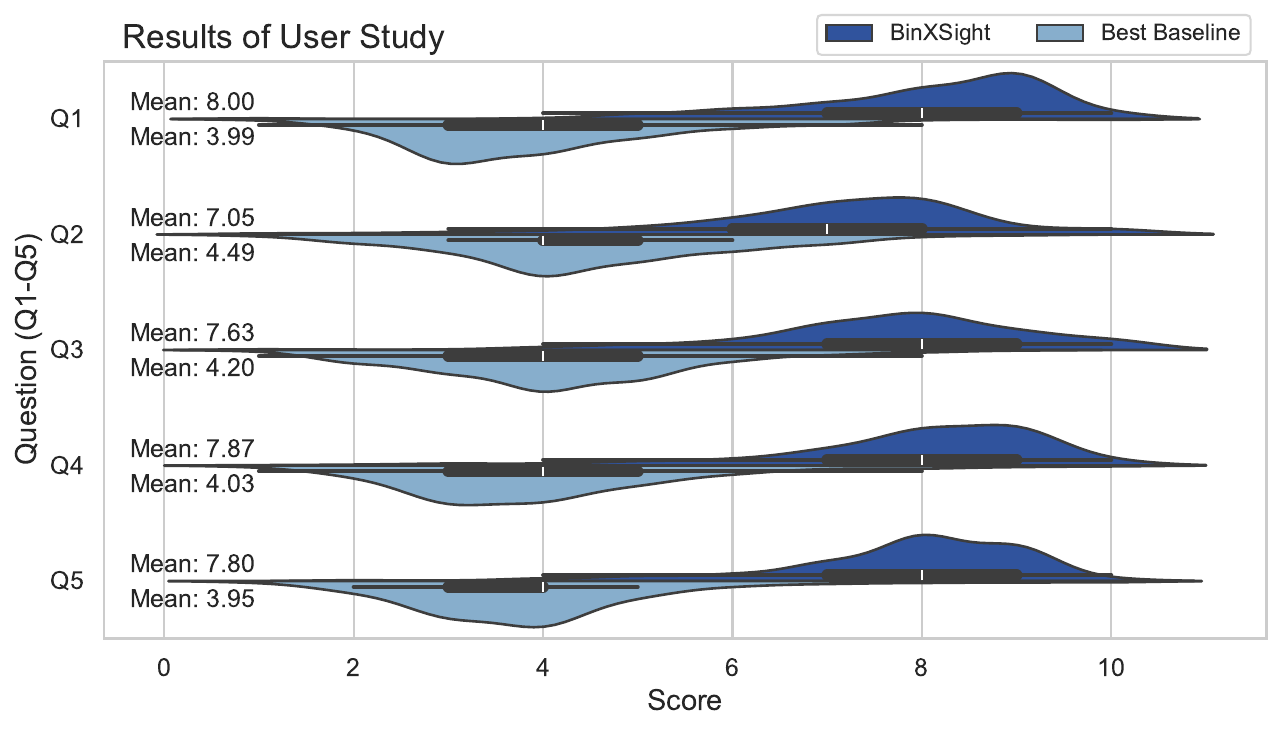}
\caption{User study results. The baseline denotes the one achieving the highest SIMILARITY on each binary among the five.}
\label{fig:questionnaire}
\end{figure}

\smallskip
\noindent \textbf{Result.} Fig. \ref{fig:questionnaire} presents the results of the human evaluation in violin and box plots. The experimental results show that \binxsight{} overwhelmingly outperforms the baseline scores across all five questions, as evidenced by the two plots, with mean scores ranging from 7.05 to 8.00. Such a result also indicates a clear positive preference for \binxsight{}'s summaries in the qualitative evaluation of the five aspects, i.e., \textit{Meaningfulness and Helpfulness}, \textit{Minimal Superfluous Description}, \textit{Readability}, \textit{Functionality Understanding}, and \textit{Completeness}, demonstrating its practical effects of assisting reverse engineers in quickly understanding unknown binaries in the first place. In contrast, the baseline has significantly lower mean scores, ranging only from 3.95 to 4.49 $<$ 5.0. \looseness=-1

In addition, among all five questions, we observe that Q2 possesses lower mean scores than the other questions for \binxsight{}. After interviewing, we learn that it is mainly because the summaries of some binaries can contain irrelevant information such as memory management or error handling, affecting reading and understanding. This information is inherently carried by the function internal semantics (e.g., \texttt{calloc} invocation). Though we can ask LLM actively to eliminate them through a secondary prompting on the final summary, we believe it may also impact binaries that indeed possess such behaviors (e.g., glibc), affecting results.

\subsection{RQ4: Ablation Study \& Component Analysis}\label{sec:ablationstudyandcomponentanalysis}

\noindent \textbf{Motivation.} To evaluate the effectiveness of \binxsight{}’s components, we perform
a set of ablation studies and component analyses.

\smallskip
\noindent \textbf{Approach.} We validate the effectiveness of \binxsight{}'s component from two perspectives: removal (i.e., ablation study) and replacement (i.e., component analysis), on the test dataset.

\noindent \textit{Ablation Study:} We remove components with the following:

\begin{itemize}[leftmargin = *]
    \item \textbf{\binxsight{}-w/o-BG:} Remove the binary graph component. That is, only ranking and filtering are applied to func-sum and func-code. It is used to evaluate the necessity of modules.

    \item \textbf{\binxsight{}-w-FCG:} Remove FSG, LLG, and DRG to understand how helpful the three graphs are for completing binary graph.

    \item \textbf{\binxsight{}-w/o-Graph:} Remove each graph from FCG, FSG, LLG, and DRG in graph construction, evaluating their contributions to binary graph construction.

    \item \textbf{\binxsight{}-w/o-HR:} Take only level 1 modules when decomposition (functions beyond token limits are discarded) to evaluate the necessity of the hierarchical strategy.

    \item \textbf{\binxsight{}-w/o-Ra:} Remove the \textbf{\textit{Ranking}} to evaluate its necessity of selecting high-value feature-optimized modules.

    \item \textbf{\binxsight{}-w/o-Re:} Remove the \textbf{\textit{Reconstruction}} to evaluate the effectiveness of the hierarchy information of modules.

    \item \textbf{\binxsight{}-w/o-FNP:} Remove the function name prediction to assess how helpful the hints provided are.

    \item \textbf{\binxsight{}-w/o-Feature:} Remove ranking features of LoC and PoU in a cumulative manner to evaluate their contributions to our \textbf{Ranking} strategy.
\end{itemize}

\noindent \textit{Component Analysis:} We replace components with following:

\begin{itemize}[leftmargin = *]

    \item \textbf{\binxsight{}-r-FS:} Replace our \textbf{Leaf-node Module Summarization} with function summarization first and then for module. It can evaluate the effectiveness of our module summarization.

    \item \textbf{\binxsight{}-r-DeLink:} Replace our graph construction and modular decomposition with the state-of-the-art DeLink~\cite{lang2024delink}. Note that it is designed to recover only compile units.

    \item \textbf{\binxsight{}-r-ModX:} Replace our graph construction and modular decomposition with the state-of-the-art ModX~\cite{yang2022modx}. Note that it is designed to recover TPL structures.

    \item \textbf{\binxsight{}-r-CWS:} Replace $S_p$ with different context window sizes. It is used to evaluate the effectiveness of \textbf{\textit{Filtering}} and the influence of different $S_p$.

    \item \textbf{\binxsight{}-r-LLM:} Replace the base LLM with kimi-k2.5~\cite{team2026kimi} and claude-opus-4-6~\cite{claude-opus-4-6} to explore the differences between LLMs. However, the study of the kind of LLM that can better drive \binxsight{} is not the focus of our work.

\end{itemize}

\begin{table}[t]
    \centering
    \caption{The ablation study results for \binxsight{}. Each component provides a positive contribution.}
    \label{tab:ablationstudy}
    \scalebox{0.75}{
    \begin{tabular}{c|c|c|c|c}
    \toprule
    \textbf{Method} & \textbf{SIMILARITY} &\textbf{BLEU} & \textbf{ROUGE-L} & \textbf{METEOR} \\
    \midrule

    \binxsight{}-w/o-BG-func-sum     & 0.4938 & 3.2242 & 0.1514 & 0.2230 \\
    \binxsight{}-w/o-BG-func-code    & 0.5262 & 3.7762 & 0.1513 & 0.2199 \\

    \binxsight{}-w-FCG   & 0.5623 & 4.0168 & 0.1697 & 0.2435 \\

    \binxsight{}-w/o-HR  & 0.5675 & 3.9527 & 0.1701 & 0.2409 \\

    \binxsight{}-w/o-Ra  & 0.5693 & 3.9475 & 0.1682 & 0.2403 \\
 
    \binxsight{}-w/o-Re  & 0.5746 & 3.9124 & 0.1684 & 0.2423 \\

    \binxsight{}-w/o-FNP  & 0.5805 & 3.9852 & 0.1621 & 0.2355 \\

    \binxsight{} & \textbf{0.6137} & \textbf{4.6204}   & \textbf{0.1897}   & \textbf{0.2548} \\

    \bottomrule
    \end{tabular}
    }
\end{table}

\begin{figure}[t]
    \centering
    \subfigure[SIMILARITY and BLEU across variants in graph.]{
    \includegraphics[width=0.48\textwidth]{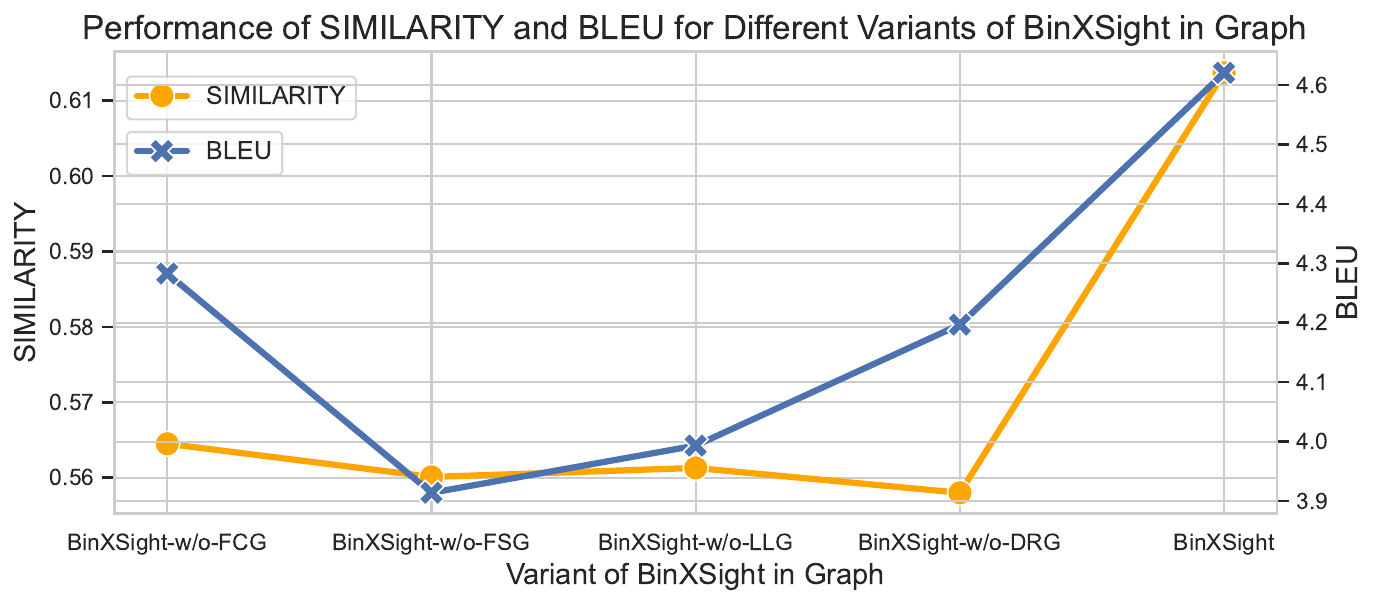}}
    \subfigure[ROUGE-L and METEOR across variants in graph.]{
    \includegraphics[width=0.48\textwidth]{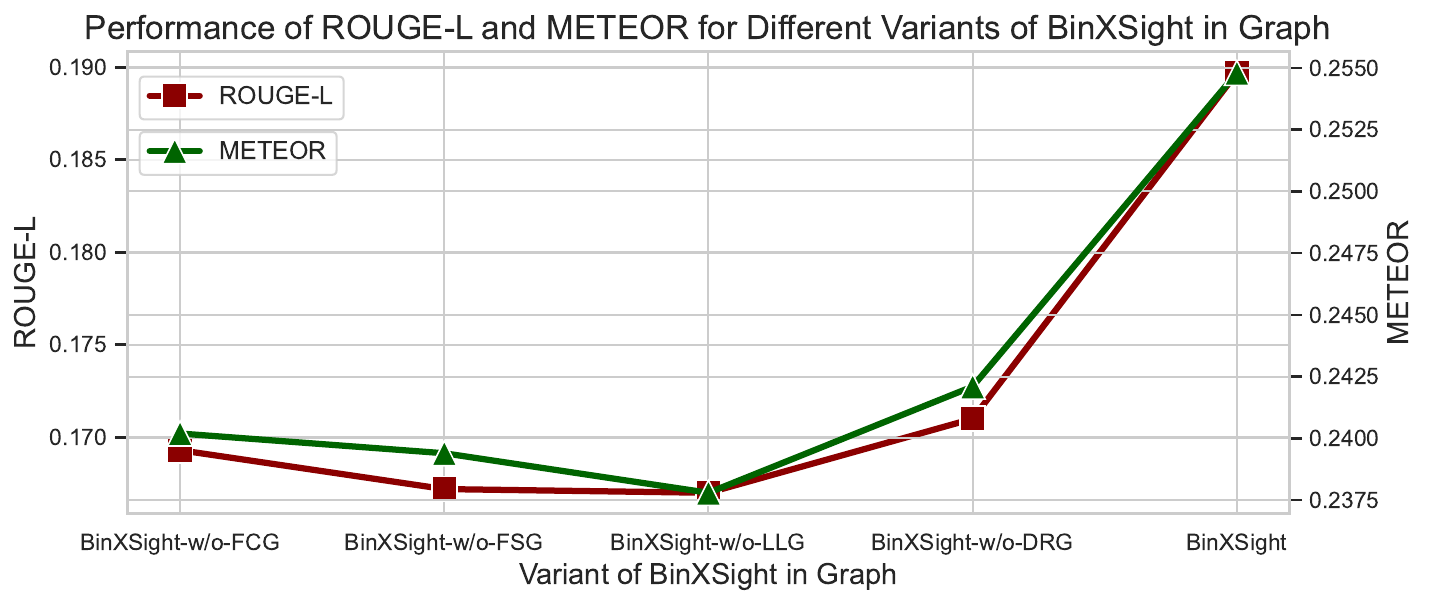}}
    \caption{Metric values across variants of \binxsight{} in graph.}
    \label{fig:graph}
\end{figure}

\begin{figure}[t]
    \centering
    \subfigure[SIMILARITY and BLEU across variants of \binxsight{} in Ranking.]{
    \includegraphics[width=0.48\textwidth]{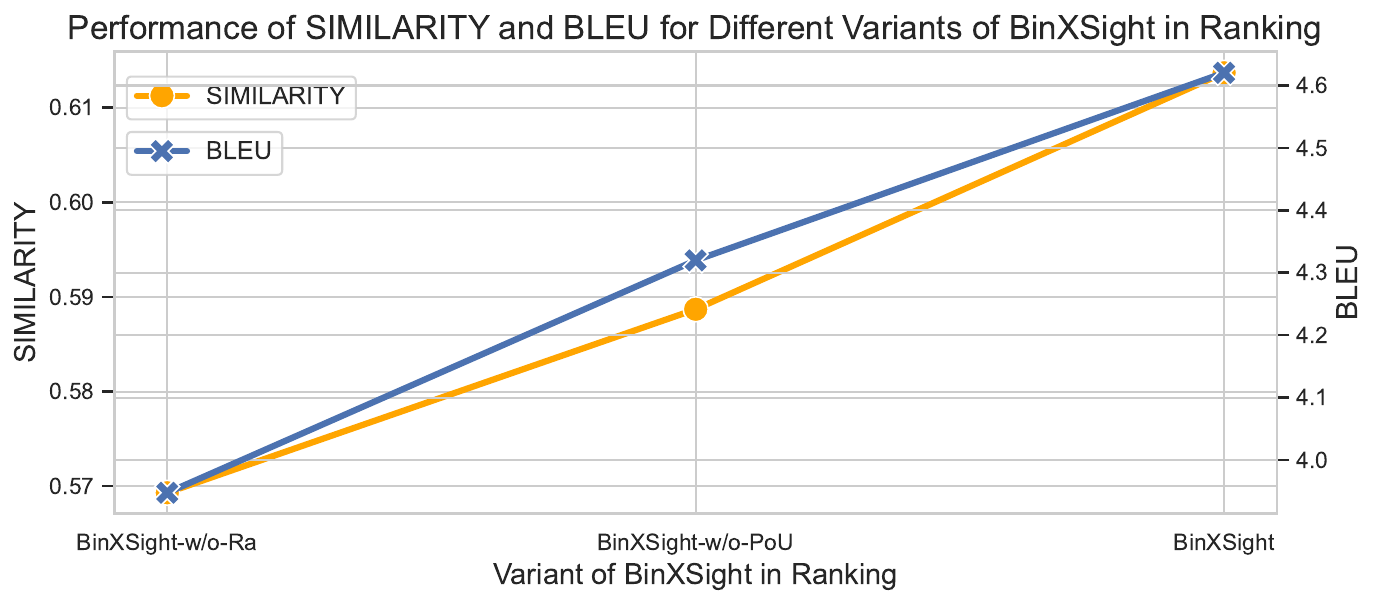}}
    \subfigure[ROUGE-L and METEOR across variants of \binxsight{} in Ranking.]{
    \includegraphics[width=0.48\textwidth]{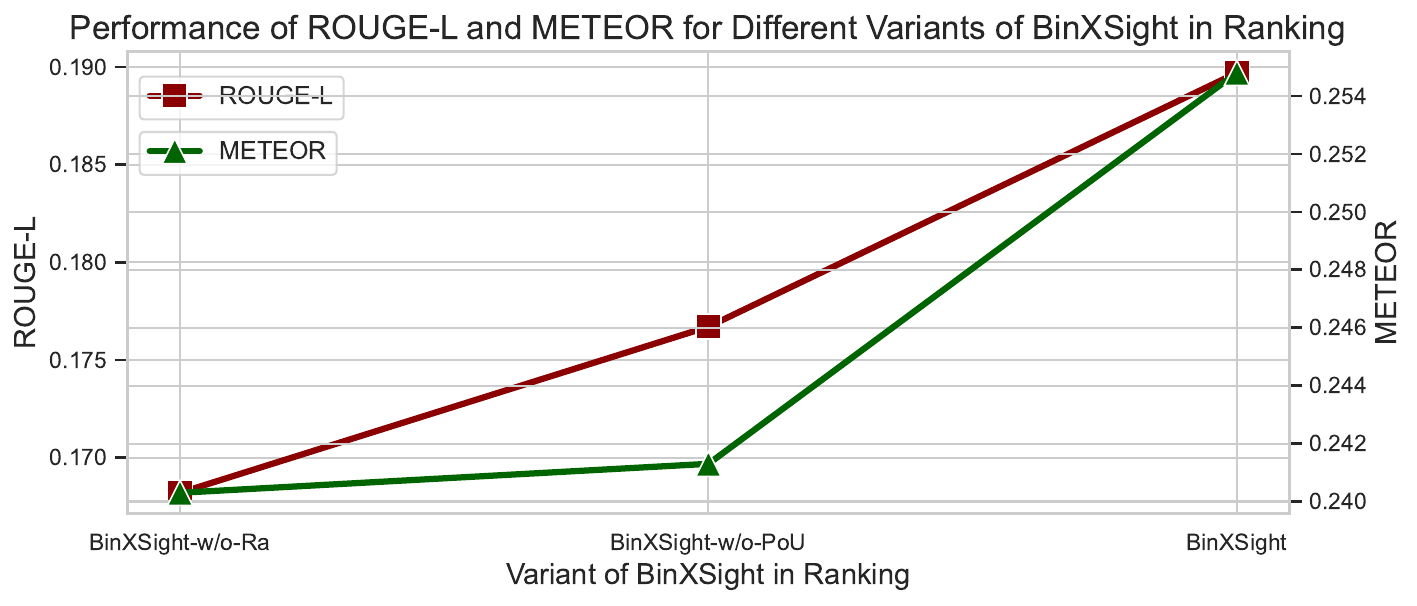}}
    \caption{Metric values across variants of \binxsight{} in Ranking.}
    \label{fig:ra}
\end{figure}

\smallskip
\noindent \textbf{Result.} We first introduce the \textit{ablation study} results.
As shown in Table \ref{tab:ablationstudy}, Fig. \ref{fig:graph}, and Fig. \ref{fig:ra}, each component provides a positive contribution. Among these, binary graph with modular decomposition has the most significant impact. Both \binxsight{}-w/o-BG achieve the lowest on the four metrics.
\looseness=-1

For \binxsight{}-w-FCG, it achieves lower 0.5623, 4.0168, 0.1697, and 0.2435 in the four metrics. Relying solely on FCG to generate modules can cause the loss of some important functions that contribute to the same program behavior. LLM cannot see such functions when summarizing these incomplete modules, resulting in negative impacts. 
\looseness=-1

For \binxsight{}-w/o-Graph (Fig. \ref{fig:graph}), the four variants of removing each graph from the binary graph achieve a lower performance compared with \binxsight{}. These results demonstrate that each graph from FCG, FSG, LLG, and DRG has a positive contribution to the binary graph for program summarization.
\looseness=-1

Regarding \binxsight{}-w/o-HR, \binxsight{} outperforms it in the four metrics, illustrating the necessity of the hierarchical strategy when decomposition. 
\looseness=-1

Compared with \binxsight{}-w/o-Ra and Re, \binxsight{} outperforms them by at least 6.80\%, 17.05\%, 12.65\%, and 5.16\% in the four metrics. It demonstrates the necessity of selecting high-value modules and reconstructing the hierarchy for module summaries, particularly for binaries with numerous functions.
\looseness=-1

As for \binxsight{}-w/o-FNP, predicting function names as hints can also improve the performance of \binxsight{}. For example, llama2.c~\cite{llamarun} is a binary running llama LLMs. The FNP predicts a name of \code{softmaxTransformInplace} to a function with ground truth of \code{softmax}, which helps identify that the module may be used for DL inference, improving the final program summary. Without such hints, \binxsight{}-w/o-FNP infers that this module is only used for numerical computation.
\looseness=-1

For \binxsight{}-w/o-Feature (Fig. \ref{fig:ra}), continuously supplementing ranking features generally has a positive effect (i.e., $\text{w/o-Ra} \rightarrow \text{w/o-PoU} \rightarrow \text{w/o-}\phi$). It demonstrates that both features are effective in ranking high-value feature-optimized modules.
\looseness=-1

\begin{table}[t]
    \centering
    \caption{The component analysis result for \binxsight{}.}
    \label{tab:componentanalysis}
    \scalebox{0.75}{
    \begin{tabular}{c|c|c|c|c}
    \toprule
    \textbf{Method} & \textbf{SIMILARITY} &\textbf{BLEU} & \textbf{ROUGE-L} & \textbf{METEOR} \\

    \midrule

    \binxsight{}-r-FS & 0.5181 & 3.2765 & 0.1581 & 0.2218 \\

    \midrule

    \binxsight{}-r-DeLink & 0.5636 & 3.9810 & 0.1690 & 0.2382 \\

    \binxsight{}-r-ModX & 0.5627 & 4.0198 & 0.1718 & 0.2313 \\

    \midrule

    \binxsight{}-r-kimi-k2.5         & 0.6091 & 4.5843 & 0.1890 & 0.2527  \\
    \binxsight{}-r-claude-opus-4-6 & 0.6125 & 4.5688 & 0.1865 & 0.2532 \\

    \midrule

    func-sum-r-claude-opus-4-6   & 0.4750 & 2.8831 & 0.1424 & 0.2224 \\
    func-code-r-claude-opus-4-6  & 0.5026 & 3.1653 & 0.1437 & 0.2171 \\
    sym-sum-r-claude-opus-4-6    & 0.4829 & 2.9302 & 0.1430 & 0.2143 \\
    Malsight-r-claude-opus-4-6   & 0.4917 & 3.2504 & 0.1544 & 0.2192 \\

    \midrule

    \binxsight{} & \textbf{0.6137} & \textbf{4.6204}   & \textbf{0.1897}   & \textbf{0.2548} \\

    \bottomrule
    \end{tabular}
    }
\end{table}

\begin{figure}[t]
    \centering
    \subfigure[SIMILARITY and BLEU across different CWS.]{
    \includegraphics[width=0.48\textwidth]{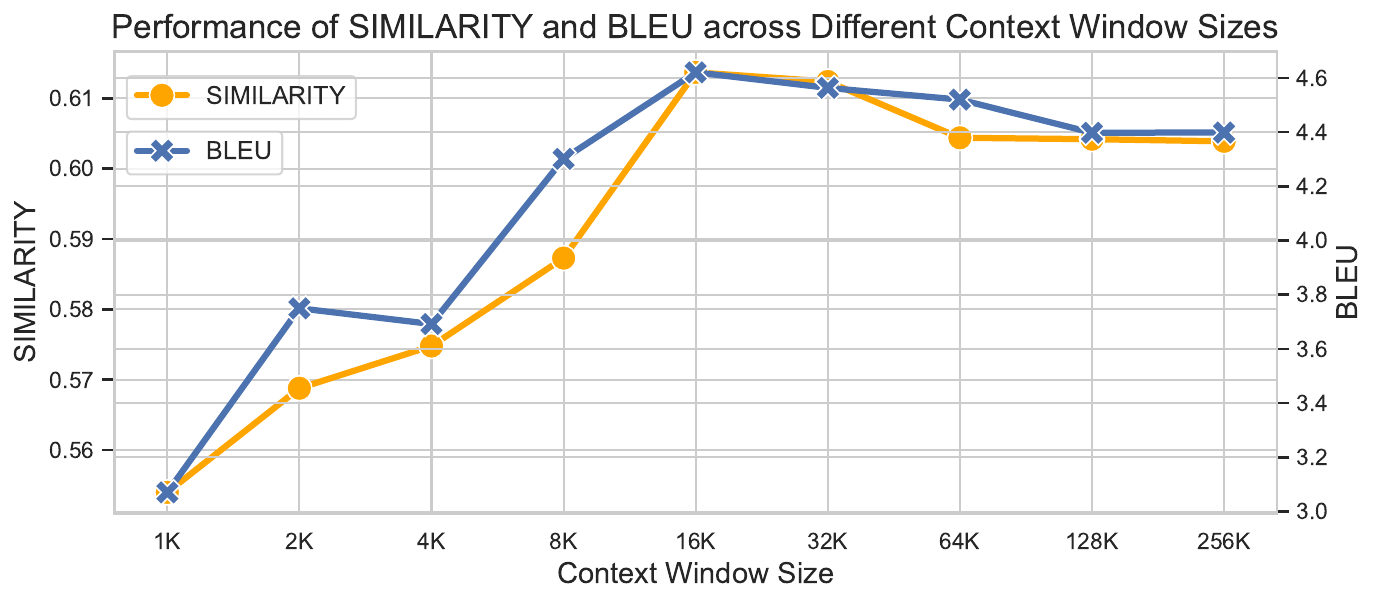}}
    \subfigure[ROUGE-L and METEOR across different CWS.]{
    \includegraphics[width=0.48\textwidth]{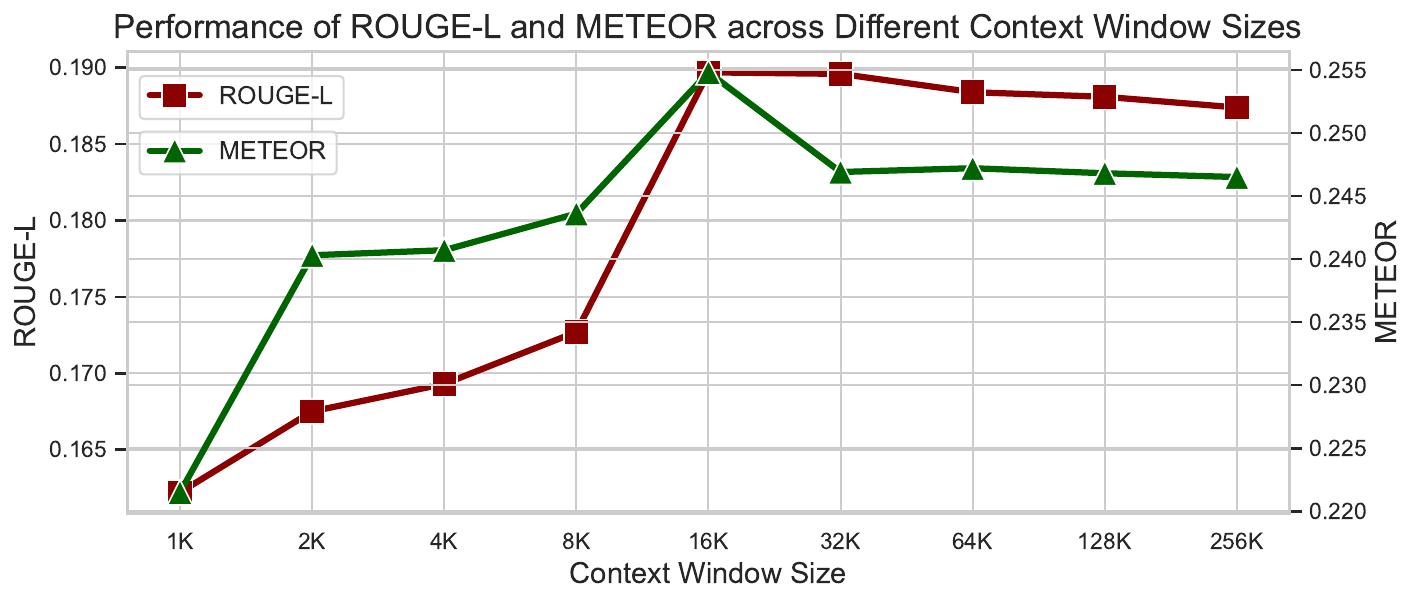}}
    \caption{Metric values across different CWS of \binxsight{}.}
    \label{fig:cws}
\end{figure}

Next, we investigate the \textit{component analysis} results.
The experimental results are shown in Table \ref{tab:componentanalysis} and Fig. \ref{fig:cws}. For \binxsight{}-r-FS, \binxsight{} significantly outperforms it by 18.45\%, 41.02\%, 19.99\%, and 14.88\% in the four metrics. The prioritized function summarization loses the context information, thereby affecting the LLM's understanding of both modules and programs.
\looseness=-1

For changing to DeLink and ModX, they achieve lower scores in the four metrics. This result indicates modules decomposed by \binxsight{} can provide better core behavioral semantics.
\looseness=-1

Moreover, the comparisons of different CWS of $S_p$ to \binxsight{} are shown in Fig. \ref{fig:cws}. The results indicate that different $S_p$ affect the performance. When $S_p$ is too small (e.g., $<$ 16K), high-value modules are filtered out, weakening the effectiveness of \binxsight{}. Conversely, when $S_p$ is too large (e.g., $>$ 16K), an excess of low-value modules is incorporated. Consequently, the performance of \binxsight{} does not improve and may even decline. An excessively large $S_p$ will also reduce efficiency and incur higher economic costs. Thus, we set $S_p =$ 16K as it has the best performance with a computationally-reasonable CWS when performing \textbf{\textit{Filtering}}. \looseness=-1

For \binxsight{}-r-LLM, different base LLMs exhibit varying performance (e.g., \binxsight{}-r-claude-opus-4-6 achieves 0.6125 in SIMILARITY). However, overall, there is no overwhelming advantage among them.
We also re-evaluate all baselines with claude-opus-4-6, and \binxsight{} still substantially outperforms them, confirming that its advantage stems from the framework design rather than the base LLM.
\looseness=-1

\subsection{Case Analysis}\label{sec:caseanalysis}

\noindent \textbf{Motivation.} We further investigate the specific cases to illustrate \binxsight{}'s practical utility in malware security.

\begin{table}[t]
    \centering
    \caption{The summaries for stripped Mirai malware by \binxsight{} and the baseline with the best SIMILARITY score.}
    \label{tab:miraicaseanalysis}
    \scalebox{0.75}{
    \begin{tabular}{p{0.6\textwidth}|p{0.6\textwidth}}
    \toprule
    \multicolumn{1}{c|}{\textbf{\binxsight{}}} & \multicolumn{1}{c}{\textbf{Baseline (Malsight)}} \\
    
    \midrule

    The binary program functions as a \textbf{\textit{\textcolor{red}{network-based attack tool that facilitates malicious operations on remote systems. It is designed to manage and execute various network flooding attacks, utilizing protocols like HTTP, TCP, and UDP.}}} The program establishes socket connections for network communication, processes incoming commands, and manages authentication entries, indicating a command and control focus. It includes features for customizing attack parameters, generating random data, and ensuring single-instance execution. Overall, the program serves as a comprehensive toolkit for executing network attacks and managing related processes, likely as part of a larger malware framework. & \textbf{The binary program is a complex system designed for managing and monitoring network connections, likely in a server or embedded environment.} It includes various functions for handling TCP connections, managing watchdog timers, and processing network data. The program utilizes a series of system calls and interrupts to interact with the operating system, allowing it to read and write data to network interfaces and manage resources efficiently. The program also implements a watchdog mechanism to ensure that processes remain responsive. It features a robust error handling system, utilizing software interrupts to manage various states and conditions. Overall, the binary serves as a foundational component for network communication, ensuring reliable data transfer and system stability. \\

    \bottomrule
    \end{tabular}
    }
\end{table}

\smallskip
\noindent \textbf{Malware Analysis.} We first give an in-the-wild example of Mirai~\cite{antonakakis2017understanding} (in the test dataset), the notorious botnet family that targets IoT devices for DDoS attacks. The Table \ref{tab:miraicaseanalysis} shows the malware's \binxsight{}-generated summary. As we can see, \binxsight{} provides an accurate summary of Mirai's DDoS (i.e., flooding) attack behaviors, which can help users quickly understand and capture the malicious functionality of it. In contrast, the baseline fails to capture the functionality and cannot provide effective assistance.

\begin{table}[t]
    \centering
    \caption{Summaries for the recent active real-world malware samples \cite{malware5} and \cite{malware55}, collected by the security vendors.}
    \label{tab:realwordcases}
    \scalebox{0.75}{
    \begin{tabular}{p{0.6\textwidth}|p{0.6\textwidth}}
    \toprule

    \multicolumn{1}{c|}{\textbf{Summary to Malware~\cite{malware5} (6/72 AV Engines)}} & \multicolumn{1}{c}{\textbf{Summary to Malware~\cite{malware55} (54/73 AV Engines)}} \\

    \midrule

    \textbf{\textit{\textcolor{red}{The binary program has various functions, including performing malicious activities by creating processes, loading libraries, moving files, and manipulating the registry.}}} It attempts to establish persistence in the system registry and may try to disable anti-spyware software. Additionally, the program seeks to connect to specific IP addresses, potentially for remote control or data transmission. It also involves moving files to the AppData directory, which may be intended to conceal its activities. & The binary program appears to be an application for the Windows platform, primarily focused on system management and network operations. It may run in the background, register itself as a startup item, and copy files to specific directories. The program includes operations on the registry, potentially to modify or add startup items. Additionally, it involves network access and may connect to specific websites. \textbf{\textit{\textcolor{red}{Overall, this program could be considered malware or a tool designed for system monitoring or data collection without the user's knowledge.}}} \\

    \bottomrule
    \end{tabular}
    }
\end{table}

To validate the capabilities of \binxsight{} in real-world security scenarios, we provide SDK to security vendors for their use in practical malware threat analysis. \textit{\textit{After a week-long wait, we received positive feedback that \binxsight{} is capable of enhancing the malware analysis process with timely insights into potential threats}.} For example, Table \ref{tab:realwordcases} presents the summarized results of two recent active real-world malware cases from security vendors' feedback. It effectively identifies their suspicious and malicious behaviors, showing practical utility in malware threat intelligence.
\looseness=-1

Despite performing well above, \binxsight{} also has failure cases. For the gh0st server~\cite{gh0st}, it devotes much of the summary to MFC framework and GUI details, while only briefly covering its malicious behaviors, such as file management, audio processing, and remote control. 
The core behaviors are still captured, but the emphasis on UI details dilutes the malicious intent. 
This is because some modules involve MFC framework functions, and although the partition is reasonable (the functions within each module are indeed grouped for the same behavior), the LLM still describes the MFC functionality.
Instructing the prompt to suppress MFC descriptions can remove such information while retaining the core malicious behaviors.
\looseness=-1

%% file: secs/discussion.tex
\section{Discussion}\label{sec:discussion}

\smallskip
\noindent \textbf{Heuristic Filtering and Robustness.} \binxsight{} is designed as a fast and effective approach, for which we adopt heuristic ranking (LoC and PoU) and context-window-based filtering to retain only high-value modules.
We acknowledge that such heuristics may discard a few useful modules or retain unnecessary ones, such as the GUI/MFC modules discussed earlier.
Nevertheless, we argue this is an acceptable and efficient trade-off.
As our experiments show, \binxsight{} remains effective across binaries of diverse types and sizes, indicating that ranking and filtering are sound insights for capturing core behaviors at low cost.
We further evaluate a variant that retains all modules: instead of filtering, it processes module summaries in successive batches, generating an initial program summary and iteratively refining it with each subsequent batch until all modules are incorporated.
This variant achieves 0.6011, 4.3863, 0.1854, and 0.2437 in the four metrics, comparable to (slightly below) the original 0.6137, 4.6204, 0.1897, and 0.2548, while incurring higher cost (e.g., 19$\times$ for the largest binary Z3). Thus, using all modules is available as an optional setting for robustness, but our filtering design attains comparable quality far more efficiently.
Looking forward, a more semantically-aware, dynamic ranking-filtering strategy is a promising direction to further mitigate potential information loss.
\looseness=-1

\begin{table}[!t]
    \centering
    \caption{The performance of \binxsight{} across optimization levels, architectures, and obfuscations.}
    \label{tab:compileenvironment}
    \resizebox{\textwidth}{!}{
    \begin{tabular}{c|c|cccc|cccc|cccc}
    \toprule
    \multirow{2}{*}{\textbf{Method}} & \multirow{2}{*}{\textbf{Metric}} & \multicolumn{4}{c|}{\textbf{Optimization Level}} & \multicolumn{4}{c|}{\textbf{Architecture}} & \multicolumn{4}{c}{\textbf{Obfuscation}} \\
    & & O0 & O1 & O2 & O3 & x86 & x64 & ARM & MIPS & SUB & BCF & FLA & ALL \\
    \midrule
    \multirow{4}{*}{func-sum} & SIMILARITY & 0.4772 & 0.4509 & 0.4697 & 0.4476 & 0.4842 & 0.4850 & 0.4857 & 0.4633 & 0.4488 & 0.4291 & 0.4270 & 0.4134 \\
     & BLEU & 3.5254 & 3.2435 & 3.4943 & 3.6016 & 3.5419 & 3.4776 & 3.8804 & 3.3297 & 3.3644 & 2.9739 & 2.8092 & 2.7030 \\
     & ROUGE-L & 0.1533 & 0.1367 & 0.1412 & 0.1421 & 0.1493 & 0.1472 & 0.1499 & 0.1464 & 0.1322 & 0.1355 & 0.1274 & 0.1200 \\
     & METEOR & 0.2467 & 0.2370 & 0.2517 & 0.2385 & 0.2412 & 0.2353 & 0.2268 & 0.2329 & 0.2261 & 0.2143 & 0.2166 & 0.2052 \\
    \midrule
    \multirow{4}{*}{func-sum-MiSum} & SIMILARITY & 0.4258 & 0.4114 & 0.4201 & 0.4046 & 0.4314 & 0.4441 & 0.4252 & 0.4130 & 0.4058 & 0.3886 & 0.3756 & 0.3472 \\
     & BLEU & 2.8610 & 2.5085 & 2.5848 & 2.7206 & 2.4595 & 2.6411 & 2.7422 & 2.7892 & 2.1285 & 2.3994 & 2.0569 & 1.8981 \\
     & ROUGE-L & 0.1391 & 0.1228 & 0.1345 & 0.1319 & 0.1271 & 0.1318 & 0.1290 & 0.1377 & 0.1152 & 0.1203 & 0.1176 & 0.1068 \\
     & METEOR & 0.2104 & 0.2018 & 0.2271 & 0.2095 & 0.2122 & 0.2065 & 0.1970 & 0.2013 & 0.1925 & 0.2131 & 0.1893 & 0.1804 \\
    \midrule
    \multirow{4}{*}{func-code} & SIMILARITY & 0.5228 & 0.4950 & 0.5189 & 0.5047 & 0.4857 & 0.5110 & 0.5025 & 0.5062 & 0.4692 & 0.4741 & 0.4494 & 0.4261 \\
     & BLEU & 4.0305 & 3.7690 & 3.8978 & 3.9007 & 3.9967 & 3.7331 & 4.2840 & 3.9113 & 3.5774 & 3.2573 & 3.1606 & 3.0412 \\
     & ROUGE-L & 0.1523 & 0.1401 & 0.1505 & 0.1490 & 0.1426 & 0.1463 & 0.1432 & 0.1495 & 0.1364 & 0.1345 & 0.1287 & 0.1248 \\
     & METEOR & 0.2396 & 0.2350 & 0.2321 & 0.2307 & 0.2250 & 0.2421 & 0.2295 & 0.2252 & 0.2276 & 0.2143 & 0.2148 & 0.1984 \\
    \midrule
    \multirow{4}{*}{sym-sum} & SIMILARITY & 0.4518 & 0.4650 & 0.4781 & 0.4779 & 0.4848 & 0.4943 & 0.4662 & 0.4574 & 0.4465 & 0.4447 & 0.4169 & 0.3892 \\
     & BLEU & 3.4212 & 3.6691 & 3.4157 & 3.5469 & 3.6914 & 3.7453 & 3.5710 & 3.5192 & 3.4805 & 2.8437 & 2.8174 & 2.6386 \\
     & ROUGE-L & 0.1321 & 0.1430 & 0.1527 & 0.1474 & 0.1389 & 0.1502 & 0.1503 & 0.1445 & 0.1257 & 0.1289 & 0.1273 & 0.1190 \\
     & METEOR & 0.2308 & 0.2377 & 0.2486 & 0.2180 & 0.2222 & 0.2329 & 0.2318 & 0.2265 & 0.2181 & 0.2104 & 0.2143 & 0.2012 \\
    \midrule
    \multirow{4}{*}{Malsight} & SIMILARITY & 0.4796 & 0.4923 & 0.4907 & 0.4906 & 0.4902 & 0.4811 & 0.4967 & 0.5013 & 0.4756 & 0.4686 & 0.4574 & 0.4398 \\
     & BLEU & 4.1308 & 3.6954 & 3.8761 & 3.9592 & 4.0624 & 3.9899 & 4.2217 & 3.9659 & 3.6965 & 3.2900 & 3.2836 & 3.0212 \\
     & ROUGE-L & 0.1604 & 0.1508 & 0.1551 & 0.1573 & 0.1607 & 0.1585 & 0.1579 & 0.1574 & 0.1393 & 0.1418 & 0.1391 & 0.1290 \\
     & METEOR & 0.2339 & 0.2373 & 0.2371 & 0.2315 & 0.2360 & 0.2443 & 0.2264 & 0.2299 & 0.2362 & 0.2144 & 0.2226 & 0.2013 \\
    \midrule
    \multirow{4}{*}{\binxsight{}} & SIMILARITY & \textbf{0.6091} & \textbf{0.6050} & \textbf{0.6173} & \textbf{0.6044} & \textbf{0.6133} & \textbf{0.6173} & \textbf{0.6078} & \textbf{0.6184} & \textbf{0.5829} & \textbf{0.5755} & \textbf{0.5638} & \textbf{0.5718} \\
     & BLEU & \textbf{5.9513} & \textbf{5.3980} & \textbf{5.6081} & \textbf{5.8286} & \textbf{5.8842} & \textbf{5.6081} & \textbf{6.1501} & \textbf{5.6297} & \textbf{5.3338} & \textbf{4.7548} & \textbf{4.6216} & \textbf{4.6057} \\
     & ROUGE-L & \textbf{0.1990} & \textbf{0.1856} & \textbf{0.1964} & \textbf{0.1977} & \textbf{0.1954} & \textbf{0.1964} & \textbf{0.1960} & \textbf{0.1934} & \textbf{0.1773} & \textbf{0.1751} & \textbf{0.1730} & \textbf{0.1663} \\
     & METEOR & \textbf{0.2757} & \textbf{0.2693} & \textbf{0.2806} & \textbf{0.2719} & \textbf{0.2722} & \textbf{0.2806} & \textbf{0.2659} & \textbf{0.2715} & \textbf{0.2685} & \textbf{0.2516} & \textbf{0.2567} & \textbf{0.2514} \\
    \bottomrule
    \end{tabular}
    }
\end{table}

\smallskip
\noindent \textbf{Influence of Compile Environments.} We explore how compile environments can affect \binxsight{}. In detail, we evaluate it from three aspects: \whiteding{1} optimization levels, \whiteding{2} architectures, and \whiteding{3} obfuscations.
We first randomly select 40 compilable projects from our test dataset. Then, for \whiteding{1}, we compile them in x64 architecture with O0 to O3. For \whiteding{2}, we compile these projects in the default O2 across x86, x64, ARM, and MIPS. As for \whiteding{3}, we leverage Obfuscator-LLVM~\cite{ieeespro2015-JunodRWM} to obfuscate them (in O2 and x64) with four obfuscation options, following the prior works~\cite{keenhash:2025, benoit2023scalable, symgen:2025, jin2022symlm}: instruction substitution (SUB), bogus control flow (BCF), control flow flattening (FLA), and all combined (ALL).
SUB transforms fragments of assembly code to their equivalent form through predefined rules; BCF modifies the control flow graph (CFG) of functions by adding a large number of irrelevant basic blocks and branches; FLA changes the original CFG using a complex hierarchy of new conditions as switches; and ALL combines all obfuscations above~\cite{ieeespro2015-JunodRWM}.
\looseness=-1

The evaluation results are shown in Table~\ref{tab:compileenvironment}, where \binxsight{} consistently outperforms all five baselines.
For \whiteding{1} and \whiteding{2}, the four metrics have advantages and disadvantages as the optimization level and architecture change. While the two factors do not lead to significant differences in the results, reflecting the stability of \binxsight{} facing them.
When it comes to obfuscations (i.e., \whiteding{3}), with the SUB option, all four metrics decrease, and the decline is more significant for BCF, FLA, and ALL. The obfuscation impairs LLM's ability to comprehend code, affecting the final program summarization performance (e.g., some functionalities are incomplete), even for SUB with relatively weaker obfuscation capabilities.
Nevertheless, \binxsight{} remains more robust than all baselines under every obfuscation setting.
In this study, how to resolve obfuscation is out of our scope, and we consider this issue can be improved by applying corresponding deobfuscation tools or with the advancements of LLM. \looseness=-1

\smallskip
\noindent \textbf{Split between Benign and Malicious Binaries.} We present the experimental results of \binxsight{} on benign and malicious binaries, respectively, to demonstrate its effectiveness on both categories. \binxsight{} achieves 0.6169, 4.7031, 0.1908, and 0.2570 for benign binaries, and 0.5984, 4.4387, 0.1845, and 0.2449 for malicious binaries, in the four metrics. Such a result illustrates that \binxsight{} achieves comparable performance on both of them.

\smallskip
\noindent \textbf{Comparison of Binary Graph to Previous Works.} It is worth noting that our binary graph is different from the constructed ones from previous works~\cite{lang2024delink, karande2018bcd}, such as the state-of-the-art DeLink~\cite{lang2024delink} and ModX~\cite{yang2022modx}, because: (1) binary graph is more comprehensive than theirs. For instance, FSG (based on function semantics) is able to connect functions with similar behavior for revealing implicit behavioral relationships while they cannot; and, (2) binary graph is a representation for complete functional module generation rather than their recovering compile units~\cite{lang2024delink} or targeting on TPL detection~\cite{yang2022modx} (e.g., incomplete program behavior. See the example in LLG). It is more suitable for program summarization (see Sec. \ref{sec:ablationstudyandcomponentanalysis}).

\smallskip
\noindent \textbf{Function Name Prediction.}
Recalling \textbf{Leaf-node Module Summarization} in Sec. \ref{sec:modularoptimizationandsummarization}, to help LLM better comprehend the semantics of functions, we perform function name prediction (FNP) to provide hints. Existing generative-based methods, such as the state-of-the-art LLM-powered SymGen~\cite{symgen:2025}, depend on their understanding of function semantics. The impact of compile environments on semantic understanding can lead to their prediction of less effective names~\cite{symgen:2025, jin2023binary}, diminishing the performance in practical usage.
However, aligning binary and source functions~\cite{keenhash:2025} into the same space, the relationships of their semantics are presented in the form of the similarity score, which means that the deviation of semantics is quantified. Thus, as long as semantically similar functions maintain a certain proximity in the space, they are unaffected by the comprehension bias (e.g., compile environments). Inspired by it, we leverage function clone search~\cite{wang2024clap, wang2022jtrans} (take 0.75 as the threshold) in a binary-to-source manner to predict function names. Specifically, we choose the state-of-the-art KEENHash function embedding model~\cite{keenhash:2025} to represent both pseudo and source functions. We also build a repository having only source functions for prediction. 
Overall, the search-based one can provide better performance in practice. The search-based method achieves 58.72\%, 49.63\%, and 52.14\% in Precision, Recall, and F1-Score (using the same evaluation metrics in SymGen~\cite{symgen:2025}), surpassing SymGen's 34.38\%, 33.63\%, and 33.30\%, on our test dataset of stripped binaries (see Sec. \ref{sec:dataset}).
Note that, with the 0.75 threshold, the search-based method covers 29.64\% of the functions per binary on average, ranging from 21.85\% to 37.81\%.
\looseness=-1

\smallskip
\noindent \textbf{Economic Costs.} Excessive spending on program summaries may lead to a mismatch between costs and benefits. \binxsight{} only requires a small amount of costs to summarize a binary with gpt-5.4 as the base. For the entire test dataset, \binxsight{} consumes a total of 78.1\$ with the maximum cost of 8.9\$ to one binary, which is cost-efficient.
\looseness=-1

\smallskip
\noindent \textbf{Limitations.} \binxsight{} faces limitations inherent to static analysis, particularly with packed binaries and those containing payloads, which pose challenges for accurate function extraction~\cite{mantovani2020prevalence}. Accordingly, \binxsight{} primarily targets unpacked and reasonably decompilable binaries. To handle these issues, relevant resolution techniques (e.g., sandbox) can be integrated as a preprocessing pipeline.
In addition, in this study, we employ Ghidra for decompilation. It is essential to acknowledge that such decompiler outputs may include errors inherent to the decompilation process. The outputs from different decompilers can also vary, potentially introducing biases. Despite \binxsight{}’s design being decompiler-agnostic, exploring it with outputs from alternative decompilers presents a promising direction for future research.
Furthermore, as discussed before, obfuscation can also impact \binxsight{}'s performance, while any deobfuscation techniques or advancements in LLM can be complementary to our approach.

%% file: secs/conclusion.tex
\section{Conclusion}\label{sec:conclusion}

In this paper, we propose \binxsight{}, a novel LLM-powered framework designed to quickly generate comprehensive program-level summaries for stripped binaries. \binxsight{} decomposes a binary into smaller functional modules hierarchically and extracts the functionality of the program from them in a MapReduce manner.
Extensive evaluations across diverse stripped binaries, together with a user study and real-world malware case analysis, demonstrate the effectiveness, efficiency, and practicality of \binxsight{}.
Our artifact is provided at: \cite{binxsight}.
\looseness=-1